\documentclass[prl,superscriptaddress,twocolumn,nofootinbib,showpacs,preprintnumbers,floatfix]{revtex4-2}

\usepackage{mathrsfs}
\usepackage{amsfonts,amssymb,amsmath}
\usepackage{graphicx,epsfig}
\usepackage{multirow}
\usepackage{verbatim}
\usepackage{color}

\usepackage{slashed}

\declareslashed{}{/}{0}{-0.05}{E}
\declareslashed{}{/}{0}{-0.05}{P}
\declareslashed{}{/}{0}{-0.10}{\Widehat{P}}

\def\fsu5{$\cal{F}$-$SU(5)$}
\def\bfsu5{$\boldsymbol{\mathcal{F}}$-$\boldsymbol{SU(5)}$}

\def\neu2{$\widetilde{\chi}_2^0$}

\def\m12{$M_{1/2}$}
\def\m3half{$M_{3/2}$}
\def\m32{$M_{32}$}

\def\bs0{$B_S^0 \rightarrow \mu^+ \mu^-$}

\def\bea{\begin{eqnarray}}
\def\eea{\end{eqnarray}}
\def\nnb{\nonumber}

\begin{document}

\title{Resolving the $(g-2)_{\mu}$ Discrepancy with $\cal{F}$-$SU$(5) Intersecting D-branes}

\author{Joseph L. Lamborn}

\affiliation{Department of Chemistry and Physics, Louisiana State University, Shreveport, Louisiana 71115 USA}

\author{Tianjun Li}

\affiliation{CAS Key Laboratory of Theoretical Physics, Institute of Theoretical Physics, 
Chinese Academy of Sciences, Beijing 100190, P. R. China}

\affiliation{ School of Physical Sciences, University of Chinese Academy of Sciences, 
No.19A Yuquan Road, Beijing 100049, P. R. China}

\author{James A. Maxin}

\affiliation{Department of Chemistry and Physics, Louisiana State University, Shreveport, Louisiana 71115 USA}

\author{Dimitri V. Nanopoulos}

\affiliation{George P. and Cynthia W. Mitchell Institute for Fundamental Physics and Astronomy, Texas A$\&$M University, College Station, TX 77843, USA}

\affiliation{Astroparticle Physics Group, Houston Advanced Research Center (HARC), Mitchell Campus, Woodlands, TX 77381, USA}

\affiliation{Academy of Athens, Division of Natural Sciences, 28 Panepistimiou Avenue, Athens 10679, Greece}


\begin{abstract}

A discrepancy between the measured anomalous magnetic moment of the muon $(g - 2)_{\mu}$ and computed Standard Model value now stands at a combined $4.2\sigma$ following experiments at Brookhaven National Lab (BNL) and the Fermi National Accelerator Laboratory (FNAL). A solution to the disagreement is uncovered in flipped $SU(5)$ with additional TeV-Scale vector-like $\bf 10 + \overline{10}$ multiplets and charged singlet derived from local F-Theory, collectively referred to as $\cal{F}$-$SU(5)$. Here we engage general No-Scale supersymmetry (SUSY) breaking in $\cal{F}$-$SU(5)$ D-brane model building to alleviate the $(g - 2)_{\mu}$ tension between the Standard Model and observations. A robust $\Delta a_{\mu}$(SUSY) is realized via mixing of $M_5$ and $M_{1X}$ at the secondary $SU(5) \times U(1)_X$ unification scale in $\cal{F}$-$SU(5)$ emanating from $SU(5)$ breaking and $U(1)_X$ flux effects. Calculations unveil $\Delta a_{\mu}({\rm SUSY}) = 19.0 - 22.3 \times 10^{-10}$ for gluino masses of $M(\widetilde{g}) = 2.25 - 2.56$~TeV and higgsino dark matter, aptly residing within the BNL+FNAL $1\sigma$ mean. This $(g - 2)_{\mu}$ favorable region of the model space also generates the correct light Higgs boson mass and branching ratios of companion rare decay processes, and is further consistent with all LHC Run 2 constraints. Finally, we also examine the heavy SUSY Higgs boson in light of recent LHC searches for an extended Higgs sector.

\end{abstract}



\preprint{ACT-03-21, MI-HET-764}

\maketitle


\section{Introduction}

The Fermi National Accelerator Lab (FNAL) launched an experiment to update the original Brookhaven National Lab (BNL) analysis of the magnetic moment of the muon $(g - 2)_{\mu}$. The BNL measurements showed an enticing 3.7$\sigma$ deviation~\cite{bennett:2006fi} from the theoretical Standard Model calculation~\cite{Davier:2010nc,davier:2017zfy,keshavarzi:2018mgv,colangelo:2018mtw,hoferichter:2019gzf,davier:2019can,keshavarzi:2019abf,kurz:2014wya,gerardin:2019rua,Aubin:2019usy,giusti:2019hkz,melnikov:2003xd,masjuan:2017tvw,Colangelo:2017fiz,hoferichter:2018kwz,gerardin:2019vio,bijnens:2019ghy,colangelo:2019uex,pauk:2014rta,danilkin:2016hnh,jegerlehner:2017gek,knecht:2018sci,eichmann:2019bqf,roig:2019reh,Blum:2019ugy,colangelo:2014qya,Aoyama:2012wk,Aoyama:2019ryr,czarnecki:2002nt,gnendiger:2013pva}, albeit burdened with a large factor of uncertainty. The independently conducted FNAL venture sought to either confirm or exclude these initial BNL findings. Concluding the anticipation, FNAL announced recently a similar disparity with the Standard Model value, leading to a combined 4.2$\sigma$ discrepancy of $\Delta a_{\mu} = a_{\mu} ({\rm Exp}) - a_{\mu} ({\rm SM}) = 25.1 \pm 5.9 \times 10^{-10}$~\cite{Abi:2021gix}, in conjunction with a reduced uncertainty by a factor of four. Even more intriguing is the magnitude of the discrepancy, which just so happens to be similar in scale to the electroweak contribution~\cite{czarnecki:2002nt,gnendiger:2013pva} to $a_{\mu}$(SM), suggesting new physics obscured at the TeV-scale.

A natural explanation for the anomaly is supersymmetry (SUSY), certainly the most auspicious extension to the Standard Model. The muon's magnetic moment maintains the benefit of precision measurements and is rather sensitive to new physics, thus it has long been viewed as a gateway to probing SUSY. The triumphs of SUSY are numerous and well known with regards to stabilizing quantum corrections to the scalar Higgs field, gauge coupling unification, mechanism for radiatively breaking electroweak symmetry, and yielding a dark matter candidate in the form of the lightest supersymmetric particle (LSP) under R-parity. Our model we investigate here showcases intersecting D-branes, therefore a critical feature of SUSY is its fundamental presence in superstring theory.

The SUSY grand unification theory (GUT) model we study merges the realistic intersecting D6-brane model~\cite{Berkooz:1996km,Cvetic:2001tj,Cvetic:2001nr,Blumenhagen:2001te,Ibanez:2001nd,Cvetic:2002pj,Blumenhagen:2005mu,Cvetic:2004ui,Cvetic:2004nk,Chen:2005mj,Chen:2005aba,Chen:2005mm,Chen:2007px,Chen:2007zu,Maxin:2009ez} with the phenomenologically~\cite{Li:2010rz,Li:2010ws,Li:2010mi,Maxin:2011hy,Li:2011xua,Li:2011in,Li:2011gh,Li:2011ab,Li:2013naa,Li:2013mwa,Li:2016bww,Ford:2019kzv} and cosmologically~\cite{Ellis:2019jha,Ellis:2019bmm,Ellis:2019hps,Ellis:2019opr,Ellis:2020qad,Ellis:2020xmk,Ellis:2020nnp,Nanopoulos:2020nnh,Ellis:2020krl,Antoniadis:2020txn} favorable No-Scale flipped $SU(5)$. The union of these with extra vector-like matter, dubbed flippons~\cite{Li:2010mi}, is referred to as the \fsu5 D-brane model, which in aggregate reinforces deep theoretical constructions, furnishing a compelling natural GUT candidate for our universe. Serving as a viable high-energy candidate though, it must be capable of elegantly explaining any and all empirical observations, so we shall stress test the model to affirm whether or not it can indeed explain the BNL+FNAL 4.2$\sigma$ discrepancy (Spoiler Alert: It can!). But first we must review the \fsu5 D-brane model's foundation before we then take the deep dive into the calculations of $(g - 2)_{\mu}$ and corresponding phenomenology later in the paper.

\section{\bfsu5 Intersecting D-branes}

The \fsu5 literature is amply stocked with discussions of the minimal flipped $SU(5)$ model (for instance, see Refs.~\cite{Maxin:2011hy,Li:2011ab,Li:2013naa,Leggett:2014hha,Li:2016bww,DeBenedetti:2018fxa,DeBenedetti:2019hrk,Ford:2019kzv} and references therein). We shall provide here only a condensed review of the minimal flipped $SU(5)$ model~\cite{Barr:1981qv,Derendinger:1983aj,Antoniadis:1987dx}, where the gauge group $SU(5)\times U(1)_{X}$ is embedded into the $SO(10)$ model. First, the generator $U(1)_{Y'}$ in $SU(5)$ is defined as 
\bea 
T_{\rm U(1)_{Y'}}={\rm diag} \left(-\frac{1}{3}, -\frac{1}{3}, -\frac{1}{3},
 \frac{1}{2},  \frac{1}{2} \right)~,~\,
\label{u1yp}
\eea
which provides the hypercharge given by
\bea
Q_{Y} = \frac{1}{5} \left( Q_{X}-Q_{Y'} \right).
\label{ycharge}
\eea
There are three families of Standard Model fermions, and the quantum numbers under $SU(5)\times U(1)_{X}$ respectively are
\bea
F_i={\mathbf{(10, 1)}},~ {\bar f}_i={\mathbf{(\bar 5, -3)}},~
{\bar l}_i={\mathbf{(1, 5)}},
\label{smfermions}
\eea
with $i=1, 2, 3$. Relevant for our D-brane model notation, we associate $F_i$, ${\bar f}_i$ and ${\bar l}_i$ with the particle assignments
\bea
F_i=(Q_i, D^c_i, N^c_i),~{\overline f}_i=(U^c_i, L_i),~{\overline l}_i=E^c_i~,~
\label{smparticles}
\eea
where $Q_i$, $U^c_i$, $D^c_i$,  $L_i$, $E^c_i$ and $N^c_i$ are the left-handed quark doublets, right-handed up-type quarks, down-type quarks, left-handed lepton doublets, right-handed charged leptons, and neutrinos, respectively. Three Standard Model singlets $\phi_i$ can be introduced to generate heavy right-handed neutrino masses.

The GUT and electroweak gauge symmetries can now be broken, and this is accomplished by introducing two pairs of Higgs representations
\begin{eqnarray}
H&=&{\mathbf{(10, 1)}},~{\overline{H}}={\mathbf{({\overline{10}}, -1)}}, \nonumber \\
h&=&{\mathbf{(5, -2)}},~{\overline h}={\mathbf{({\bar {5}}, 2)}}.
\label{Higgse1}
\end{eqnarray}
The $H$ and $F$ multiplet states are labeled similarly, including only a ``bar'' added above the fields for ${\overline H}$. More precisely, the Higgs particles are
\bea
H=(Q_H, D_H^c, N_H^c)~,~
{\overline{H}}= ({\overline{Q}}_{\overline{H}}, {\overline{D}}^c_{\overline{H}}, 
{\overline {N}}^c_{\overline H})~,~\,
\label{Higgse2}
\eea
\bea
h=(D_h, D_h, D_h, H_d)~,~
{\overline h}=({\overline {D}}_{\overline h}, {\overline {D}}_{\overline h},
{\overline {D}}_{\overline h}, H_u)~,~\,
\label{Higgse3}
\eea
such that $H_d$ and $H_u$ are a single pair of MSSM Higgs doublets.

We introduce this GUT scale Higgs superpotential to break the $SU(5)\times U(1)_{X}$ gauge symmetry down to the Standard Model gauge symmetry: 
\bea
{\it W}_{\rm GUT}=\lambda_1 H H h + \lambda_2 {\overline H} {\overline H} {\overline
h} + \Phi ({\overline H} H-M_{\rm H}^2)~.~ 
\label{spgut} 
\eea
Consequently, there is only one F-flat and D-flat direction existing, which can certainly be rotated along the $N^c_H$ and ${\overline {N}}^c_{\overline H}$ directions. As a result, 
we obtain $<N^c_H>=<{\overline {N}}^c_{\overline H}>=M_{\rm H}$. Additionally, the supersymmetric Higgs mechanism allows the superfields $H$ and ${\overline H}$ to be consumed and thus acquire large masses, except $D_H^c$ and ${\overline {D}}^c_{\overline H}$. Moreover, the superpotential terms $ \lambda_1 H H h$ and $ \lambda_2 {\overline H} {\overline H} {\overline h}$ couple $D_H^c$ and ${\overline {D}}^c_{\overline H}$ respectively with $D_h$ and ${\overline {D}}_{\overline h}$, which forms massive eigenstates with masses $2 \lambda_1 <N_H^c>$ and $2 \lambda_2 <{\overline {N}}^c_{\overline H}>$. Accordingly, doublet-triplet splitting naturally occurs due to the missing partner mechanism~\cite{Antoniadis:1987dx}. There is only a small mixing through the $\mu$ term in the triplets $h$ and ${\overline h}$, so colored higgsino-exchange mediated proton decay remains negligible, {\it i.e.}, the dimension-5 proton decay problem is absent~\cite{Harnik:2004yp}. 

String-scale gauge coupling unification at about $10^{17}$~GeV can be realized by introducing the following vector-like particles (referred to as flippons) at the TeV scale derived from local F-theory model building~\cite{Jiang:2006hf, Jiang:2008yf, Jiang:2009za}
\begin{eqnarray}
&& XF ={\mathbf{(10, 1)}}~,~{\overline{XF}}={\mathbf{({\overline{10}}, -1)}}~,~\nnb \\
&& Xl={\mathbf{(1, -5)}}~,~{\overline{Xl}}={\mathbf{(1, 5)}}~.~\,
\end{eqnarray}
Under Standard Model gauge symmetry, the particle content resulting from decompositions of $XF$, ${\overline{XF}}$, $Xl$, and ${\overline{Xl}}$  are
\begin{eqnarray}
&& XF = (XQ, XD^c, XN^c)~,~ {\overline{XF}}=(XQ^c, XD, XN)~,~\nnb \\
&& Xl= XE~,~ {\overline{Xl}}= XE^c~.~
\end{eqnarray}
The additional vector-like particles under the $SU(3)_C \times SU(2)_L \times U(1)_Y$ gauge symmetry have the quantum numbers
\begin{eqnarray}
&& XQ={\mathbf{(3, 2, \frac{1}{6})}}~,~
XQ^c={\mathbf{({\bar 3}, 2,-\frac{1}{6})}} ~,~\\
&& XD={\mathbf{({3},1, -\frac{1}{3})}}~,~
XD^c={\mathbf{({\bar 3},  1, \frac{1}{3})}}~,~\\
&& XN={\mathbf{({1},  1, {0})}}~,~
XN^c={\mathbf{({1},  1, {0})}} ~,~\\
&& XE={\mathbf{({1},  1, {-1})}}~,~
XE^c={\mathbf{({1},  1, {1})}}~.~\,
\label{qnum}
\end{eqnarray}
The superpotential is
\bea 
{ W}_{\rm Yukawa} &=&  y_{ij}^{D}
F_i F_j h + y_{ij}^{U \nu} F_i  {\overline f}_j {\overline
h}+ y_{ij}^{E} {\overline l}_i  {\overline f}_j h  
\nnb \\ &&
+ \mu h {\overline h}
+ y_{ij}^{N} \phi_i {\overline H} F_j +M_{ij}^{\phi} \phi_i \phi_j
\nnb \\ &&
+ y_{XF} XF XF h + y_{\overline{XF}} {\overline{XF}} {\overline{XF}} {\overline h}
\nnb \\ &&
+ M_{XF} {\overline{XF}}  XF + M_{Xl} {\overline{Xl}}  Xl
~,~\,
\label{potgut}
\eea
and after the $SU(5)\times U(1)_X$ gauge symmetry is broken down to the Standard Model gauge symmetry, the superpotential presented directly above gives 
\bea 
{ W_{SSM}}&=&
y_{ij}^{D} D^c_i Q_j H_d+ y_{ji}^{U \nu} U^c_i Q_j H_u
+ y_{ij}^{E} E^c_i L_j H_d \nnb \\ &&
+  y_{ij}^{U \nu} N^c_i L_j H_u  +  \mu H_d H_u+ y_{ij}^{N} 
\langle {\overline {N}}^c_{\overline H} \rangle \phi_i N^c_j
\nnb \\ &&
+ y_{XF} XQ XD^c H_d + y_{\overline{XF}} XQ^c XD H_u
\nnb \\ &&
+M_{XF}\left(XQ^c XQ + XD^c XD\right) 
\nnb \\ &&
+ M_{Xl} XE^c  XE+M_{ij}^{\phi} \phi_i \phi_j
\nnb \\ &&
 + \cdots (\textrm{decoupled below $M_{GUT}$}). 
\label{poten1}
\eea
where $y_{ij}^{D}$, $y_{ij}^{U \nu}$, $y_{ij}^{E}$, $y_{ij}^{N}$, $y_{XF} $, and $y_{\overline{XF}}$ are Yukawa couplings, $\mu$ is the bilinear Higgs mass term, and $M_{ij}^{\phi}$, $M_{XF} $ and $M_{Xl}$ are new particle masses. The vector-like particle flippons are of course these new particles. The masses $M_{ij}^{\phi}$, $M_{XF}$ ,and $M_{Xl}$ have not been explicitly computed yet, reserving that in-depth project for the future. Regardless, a common mass decoupling scale $M_V$ for the vector-like multiplets is implemented.

\begin{table*}[htp]
  \centering
  \caption{The \fsu5 D-brane model general No-Scale SUSY breaking soft terms along with their associated mass spectra and other pertinent data for 12 benchmark spectra representative of that region of model space that can explain the BNL+FNAL 4.2$\sigma$ discrepancy. All spectra have higgsino LSP, with the higgsino percentage listed in the bottom line.}
\label{tab:spectra}
\begin{tabular}{|c||c|c|c|c|c|c|c|c|c|c|c|c|} \hline
$	M_1~({\rm GeV})	$&$2700  $&$2700 $&$2500  $&$ 2700 $&$	2900	$&$	2500	$&$	2800	$&$	3100	$&$	3400	$&$	3600	$&$	3500	$&$	3600	$ \\ \hline
$	M_5~({\rm GeV})	$&$1510  $&$1510   $&$1530  $&$1570  $&$	1600	$&$	1600	$&$	1600	$&$	1600	$&$	1610	$&$	1640	$&$	1660	$&$	1700	$ \\ \hline
$	M_{U^cL}~({\rm keV})	$&$ 10 $&$ 10  $&$ 10 $&$10  $&$	10	$&$	100	$&$	10	$&$	10	$&$	10	$&$	10	$&$	10	$&$	10	$ \\ \hline
$	M_{Q D^c N^c}~({\rm keV})	$&$ 10 $&$10   $&$ 10 $&$ 10 $&$	10	$&$	100	$&$	10	$&$	10	$&$	10	$&$	10	$&$	10	$&$	10	$ \\ \hline
$	M_{E^c}~({\rm GeV})	$&$1200  $&$ 1200  $&$ 1400 $&$ 1200 $&$	1300	$&$	1400	$&$	1200	$&$	1200	$&$	900	$&$	1000	$&$	1000	$&$	900	$ \\ \hline
$	M_{H_u}~({\rm GeV})	$&$1800  $&$1800   $&$1900  $&$1900  $&$	2100	$&$	2000	$&$	1900	$&$	2000	$&$	1900	$&$	2000	$&$	1900	$&$	1900	$ \\ \hline
$	M_{H_d}~({\rm GeV})	$&$2300  $&$ 2300  $&$2300  $&$ 2300 $&$	2400	$&$	2500	$&$	2400	$&$	2500	$&$	2400	$&$	2600	$&$	2600	$&$	2600	$ \\ \hline
$	A_{\tau}~({\rm GeV})	$&$800  $&$ 800  $&$ 1200 $&$400  $&$	1200	$&$	700	$&$	600	$&$	800	$&$	1200	$&$	1200	$&$	600	$&$	800	$ \\ \hline
$	A_t~({\rm GeV})	$&$600  $&$ 600  $&$200  $&$ 400 $&$	0	$&$	300	$&$	500	$&$	100	$&$	400	$&$	200	$&$	800	$&$	1000	$ \\ \hline
$	A_b~({\rm GeV})	$&$-4600  $&$ -4900  $&$-4900  $&$ -5200 $&$	-4300	$&$	-4500	$&$	-5200	$&$	-4900	$&$	-5000	$&$	-4700	$&$	-5000	$&$	-5000	$ \\ \hline
$	M_V~({\rm TeV})	$&$4000  $&$ 7000  $&$ 7000 $&$ 4000 $&$	4000	$&$	5500	$&$	5500	$&$	7000	$&$	8000	$&$	7000	$&$	8000	$&$	8000	$ \\ \hline
$	{\rm tan} \beta	$&$60  $&$60   $&$60  $&$ 60 $&$	60	$&$	59	$&$	59	$&$	60	$&$	60	$&$	60	$&$	60	$&$	60	$ \\ \hline
$	m_{\rm top}~({\rm GeV})	$&$173.3  $&$ 173.1  $&$172.7  $&$173.3  $&$	173.7	$&$	173.3	$&$	172.7	$&$	172.9	$&$	172.3	$&$	173.1	$&$	172.3	$&$	172.5	$ \\ \hline
$	{\bf \Delta a_{\mu}(SUSY)~(\times 10^{-10})}	$&${\bf 21.1}  $&${\bf 22.3}   $&$ {\bf 21.1} $&${\bf 20.1}  $&$	{\bf 19.3}	$&$	{\bf 19.0}	$&$	{\bf 19.4}	$&$	{\bf 20.2}	$&$	{\bf 20.6}	$&$	{\bf 19.0}	$&$	{\bf 19.5}	$&$	{\bf 19.0}	$ \\ \hline
$	{\rm Br}(b \to s \gamma)~(\times 10^{-4})	$&$3.01  $&$ 3.04  $&$2.99  $&$ 3.09 $&$	3.01	$&$	3.02	$&$	3.10	$&$	2.99	$&$	3.14	$&$	2.99	$&$	3.16	$&$	3.26	$ \\ \hline
$	{\rm Br}(B_S^0 \to \mu^+ \mu^-)~(\times 10^{-9})	$&$5.8  $&$4.9   $&$5.7  $&$6.2  $&$	6.2	$&$	4.6	$&$	4.6	$&$	5.6	$&$	5.7	$&$	5.2	$&$	4.3	$&$	4.1	$ \\ \hline
$	\sigma_{SI}^{\rm rescaled}~(\times 10^{-9} {\rm pb})	$&$8.6  $&$ 6.7  $&$6.8  $&$8.7  $&$	6.5	$&$	4.9	$&$	6.2	$&$	6.7	$&$	7.8	$&$	4.9	$&$	6.1	$&$	6.2	$ \\ \hline
$	\tau_p (p \to e^+ \pi^0)~(\times 10^{35} {\rm yrs})	$&$1.3  $&$ 1.3  $&$ 1.5 $&$ 1.4 $&$	1.0	$&$	1.4	$&$	1.5	$&$	1.1	$&$	1.0	$&$	0.9	$&$	1.0	$&$	1.0	$ \\ \hline
$	\Omega h^2	$&$ 0.0039 $&$0.0041   $&$ 0.0049 $&$0.0048  $&$	0.0043	$&$	0.0055	$&$	0.0050	$&$	0.0039	$&$	0.0030	$&$	0.0024	$&$	0.0032	$&$	0.0035	$ \\ \hline
$	M_{\widetilde{\chi}_1^0}~({\rm GeV})	$&$235  $&$201   $&$  202$&$ 217 $&$	185	$&$	207	$&$	215	$&$	197	$&$	190	$&$	210	$&$	197	$&$	185	$ \\ \hline
$	M_{\widetilde{\chi}_2^0}~({\rm GeV})	$&$ -282 $&$ -237  $&$  -232$&$-251  $&$	-215	$&$	-234	$&$	-247	$&$	-233	$&$	-234	$&$	-271	$&$	-240	$&$	-222	$ \\ \hline
$	M_{\widetilde{\chi}_1^{\pm}}~({\rm GeV})	$&$ 239 $&$ 206  $&$208  $&$ 223 $&$	191	$&$	213	$&$	220	$&$	202	$&$	195	$&$	213	$&$	201	$&$	190	$ \\ \hline
$	M_{\widetilde{\tau}_1^{\pm}}~({\rm GeV})	$&$ 412 $&$380   $&$352  $&$ 463 $&$	458	$&$	355	$&$	475	$&$	396	$&$	416	$&$	396	$&$	416	$&$	374	$ \\ \hline
$	M_{\widetilde{e}_R, \widetilde{\mu}_R}~({\rm GeV})	$&$1074  $&$ 1054  $&$1076  $&$ 1109 $&$	1155	$&$	1115	$&$	1113	$&$	1129	$&$	1133	$&$	1174	$&$	1162	$&$	1184	$ \\ \hline
$	M_{\widetilde{t}_1}~({\rm GeV})	$&$1737  $&$ 1738  $&$1710  $&$ 1784 $&$	1727	$&$	1770	$&$	1833	$&$	1753	$&$	1829	$&$	1814	$&$	1903	$&$	1968	$ \\ \hline
$	{\bf M_{\widetilde{g}}~({\rm \bf GeV})}	$&${\bf 2254}  $&${\bf 2281 } $&${\bf 2306}  $&${\bf 2341}  $&$	{\bf 2372}	$&$	{\bf 2388}	$&$	{\bf 2397}	$&$	{\bf 2406}	$&$	{\bf 2428}	$&$	{\bf 2463}	$&$	{\bf 2500}	$&$	{\bf 2554}	$ \\ \hline
$	M_{\widetilde{u}_R}~({\rm GeV})	$&$2522  $&$ 2507  $&$  2537$&$ 2619 $&$	2658	$&$	2651	$&$	2655	$&$	2650	$&$	2660	$&$	2713	$&$	2739	$&$	2798	$ \\ \hline
$	{\bf m_h}~({\rm \bf GeV})	$&$ {\bf 123.3} $&${\bf 123.3}   $&${\bf 123.6}  $&${\bf 124.0}  $&$	{\bf 124.6}	$&$	{\bf 124.0}	$&$	{\bf 123.6}	$&$	{\bf 124.2}	$&$	{\bf 123.4}	$&$	{\bf 124.3}	$&$	{\bf 123.1}	$&$	{\bf 123.0}	$ \\ \hline
$	M_{H^0}~({\rm GeV})	$&$936  $&$ 946  $&$ 889 $&$847  $&$	831	$&$	1014	$&$	1000	$&$	904	$&$	813	$&$	1010	$&$	962	$&$	904	$ \\ \hline
$	M_{32}~(\times 10^{16}~{\rm GeV})	$&$ 1.0 $&$1.0   $&$ 1.1 $&$1.1  $&$	1.0	$&$	1.1	$&$	1.1	$&$	1.0	$&$	1.0	$&$	0.9	$&$	1.0	$&$	1.0	$ \\ \hline
$	M_{\cal F}~(\times 10^{17}~{\rm GeV})	$&$1.9  $&$ 1.8  $&$1.8  $&$1.9  $&$	1.8	$&$	1.8	$&$	1.8	$&$	1.7	$&$	1.7	$&$	1.7	$&$	1.7	$&$	1.7	$ \\ \hline
$	{\rm LSP~Higgsino~Composition}	$&$ 67\% $&$ 80\%   $&$ 86\%  $&$ 82\%  $&$	86\% 	$&$	 89\%	$&$	 84\%	$&$	79\%	$&$	71\% 	$&$	 52\%	$&$	 72\%	$&$	 79\%	$ \\ \hline
\end{tabular}
\label{points}
\end{table*}

Contributions from vector-like multiplets require changes to the one-loop gauge $\beta$-function coefficients $b_i$ that promote a flat $SU(3)$ Renormalization Group Equation (RGE) running ($b_3 = 0$)~\cite{Li:2010ws}, separating the secondary $SU(3)_C \times SU(2)_L$ unification around $10^{16}$~GeV, which we refer to as the mass scale $M_{32}$, and the primary $SU(5) \times U(1)_X$ unification near the string scale $10^{17}$~GeV, defined as the mass scale $M_{\cal F}$. This is significant as it elevates unification close to the Planck mass. The $M3$ and $M2$ gaugino mass terms and couplings $\alpha_3$ and $\alpha_2$ unify at the scale $M_{32}$ into a single mass parameter $M5$ and coupling $\alpha_5$~\cite{Li:2010rz}, where $M5 = M3 = M2$ and $\alpha_5 = \alpha_3 = \alpha_2$ between $M_{32}$ and $M_{\cal F}$~\cite{Li:2010ws}. The flattening of the $M_3$ gaugino RGE running between $M_{32}$ and $M_V$ produces a characteristic mass texture of $M(\widetilde{t}_1) < M(\widetilde{g}) < M(\widetilde{q})$, spawning a light stop and gluino that are lighter than all other squarks~\cite{Li:2011ab}.

Two critical effects naturally transpire at the scale $M_{32}$. First, $U(1)_X$ flux effects~\cite{Li:2010rz} cause a small increase in $M_1$ from the evolution of $M_{1X}$ via the following
\begin{eqnarray}
\frac{M_1}{\alpha_1} = \frac{24}{25} \frac{M_{1X}}{\alpha_{1X}} + \frac{1}{25}\frac{M_5}{\alpha_5}
\label{flux}
\end{eqnarray}
where $\alpha_1 = 5 \alpha_Y / 3$ is the $U(1)_Y$ gauge coupling. Second, in the $SU(5)\times U(1)_X$ models motivated by D-brane model building, there exist three chiral multiplets in the $SU(5)$ adjoint representation, for example, see Ref.~\cite{Chen:2006ip}. These chiral multiplets can obtain vacuum expectation values around the $SU(3)_C \times SU(2)_L$ unification scale $M_{32}$, and then the gaugino masses for $SU(3)_C \times SU(2)_L \times U(1)_{Y'}$ of $SU(5)$ can be split due to the high-dimensional operators~\cite{Li:2010xr, Balazs:2010ha}. Because the bino mass is a linear combination of the $U(1)_{Y'}$ and $U(1)_X$ gaugino masses, the bino mass $M_1$, wino mass $M_2$, and gluino mass $M_3$ can be independent free parameters at $M_{32}$. Thus, we shall consider such effects by introducing the following relationship at the scale $M_{32}$ that stimulates mixing between $M_5$ and $M_{1X}$:
\begin{eqnarray}
M_2 = \frac{18}{25} M_5 - \frac{7}{25} M_{1X}
\label{breaking}
\end{eqnarray}
This effect drives the wino to small values at the electroweak scale and we use the resulting phenomenology to constrain the coefficients in Eq.~(\ref{breaking}). A larger contribution from $M_5$ decreases the wino contribution to $(g-2)_{\mu}$. On the contrary, a smaller contribution from $M_5$ shifts the $SU(3)_C \times SU(2)_L$ unification scale $M_{32}$ lower, to below $10^{16}$~GeV. The lower $M_{32}$ in turn pushes the proton decay rate $\tau_p (p \to e^+ \pi^0)$ down to unacceptably fast time periods of $10^{34}$~yrs or less. We study the nominal mixing highlighted in Eq.~(\ref{breaking}) in this analysis, though plan a more in-depth study later regarding maximum limits on the mixing parameters of Eq.~(\ref{breaking}).

Following upon the $F_i$, ${\bar f}_i$, and ${\bar l}_i$ of Eq.~(\ref{smparticles}), the general No-Scale SUSY breaking soft terms at $M_{\cal F}$ are $M_5$, $M_{1X}$, $M_{Q D^c N^c}$, $M_{U^c L}$, $M_{E^c}$, $M_{H_u}$,  $M_{H_d}$, $A_{\tau}$, $A_t$, and $A_b$. Note that $M_{Q D^c N^c}$ is the {\bf 10}, $M_{U^c L}$ is the $\bar{\bf 5}$, and $M_{E^c}$ is the {\bf 1} of Eq.~(\ref{smfermions}). General SUSY breaking soft terms of this type are motivated by D-brane model building~\cite{Chen:2006ip}, where $F_i$, ${\overline f}_i$, ${\overline l}_i$, and $h/{\overline h}$ arise from intersections of different stacks of D-branes. As a result, the SUSY breaking soft mass terms and trilinear $A$ terms will be different. The Yukawa terms $H H h$ and ${\overline H} {\overline H} {\overline h}$ of Eq. (\ref{spgut}) and $F_i F_j h$, $XF XF h$, and $\overline{XF} \overline{XF} {\overline h}$ of Eq.~(\ref{potgut}) are forbidden by the anomalous global $U(1)$ symmetry of $U(5)$, nonetheless, these Yukawa terms can be generated from high-dimensional operators or instanton effects. Differing from $SU(5)$ models, the Yukawa term $F_i F_j h$ in the ${\cal F}$-$SU(5)$ model gives down-type quark masses, so these Yukawa couplings can be small and hence generated via high-dimensional operators or instanton effects.

\section{Phenomenological Results}

The general No-Scale soft SUSY breaking terms in the \fsu5 D-brane model are implemented at the $SU(5) \times U(1)_X$ unification scale $M_{\cal F}$, and we concurrenly float the low-energy parameters tan$\beta$, $m_t$, and $M_V$. Over 1.2 billion points in the model space are sampled by computing the SUSY mass spectra, rare decay process branching ratios, spin-independent dark matter cross-sections, and relic density using a proprietary mpi codebase built on scaled down versions of {\tt Micromegas 2.1}~\cite{Belanger:2008sj} and {\tt SuSpect 2.34}~\cite{Djouadi:2002ze}. The intervals scanned are:
\begin{eqnarray}
\nonumber
100~{\rm GeV} \le &M_5& \le 1700~{\rm GeV} \\ \nonumber
100~{\rm GeV} \le &M_{1X}& \le 3800~{\rm GeV} \\ \nonumber
10~{\rm eV} \le &M_{U^c L}& \le 1500~{\rm GeV} \\ \nonumber
1~{\rm GeV} \le &M_{E^c}& \le 2400~{\rm GeV} \\ \nonumber
1~{\rm keV} \le &M_{Q D^c N^c}& \le 1900~{\rm GeV} \\ \nonumber
100~{\rm GeV} \le &M_{H_u}& \le 4000~{\rm GeV} \\ \nonumber
100~{\rm GeV} \le &M_{H_d}& \le 4000~{\rm GeV} \\ \nonumber
-10~{\rm TeV} \le &A_{\tau}& \le 10~{\rm TeV} \\ \nonumber
-10~{\rm TeV} \le &A_t& \le 13~{\rm TeV} \\ \nonumber
-10~{\rm TeV} \le &A_b& \le 15~{\rm TeV} \\ \nonumber
2~ \le &{\rm tan}\beta& \le 60 \\ \nonumber
1~{\rm TeV} \le &M_V& \le 8000~{\rm TeV} \\ \nonumber
\label{scan}
\end{eqnarray}

A small tolerance of $172.3 \le m_t \le 174.4$~GeV is applied around the World Average top quark mass~\cite{Olive:2016xmw}. We only consider a positive Higgs bilinear mass term $\mu$, and the strong coupling constant is fixed at $\alpha_S(M_Z) = 0.1184$, within the Particle Data Group $1\sigma$ variation~\cite{Olive:2016xmw}. The SUSY contribution to $\Delta a_{\mu}$ is calculated using {\tt GM2Calc 1.7.5}~\cite{Athron:2015rva}.

\begin{figure}[htp]
       \centering
        \includegraphics[width=0.45\textwidth]{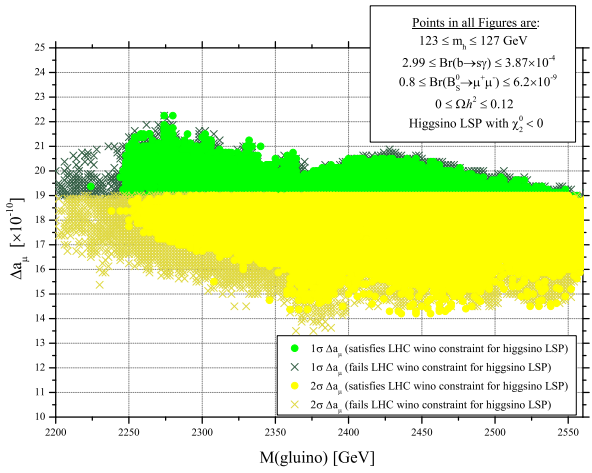}
        \caption{Depiction of $\Delta a_{\mu}$(SUSY) as a function of the gluino mass $M(\widetilde{g})$, distinguished by whether the spectra satisfy the BNL+FNAL 1$\sigma$ or $2\sigma$ limits on $\Delta a_{\mu}$. The points are also separated into groups regarding consistency with the LHC constraints on electroweakinos for those spectra with higgsino LSP, namely chargino production. The uncertainty on all $\Delta a_{\mu}$(SUSY) calculations is about $\pm 2.4 \times 10^{-10}$.}
        \label{fig:gluino}
\end{figure}

\begin{figure}[htp]
       \centering
        \includegraphics[width=0.5\textwidth]{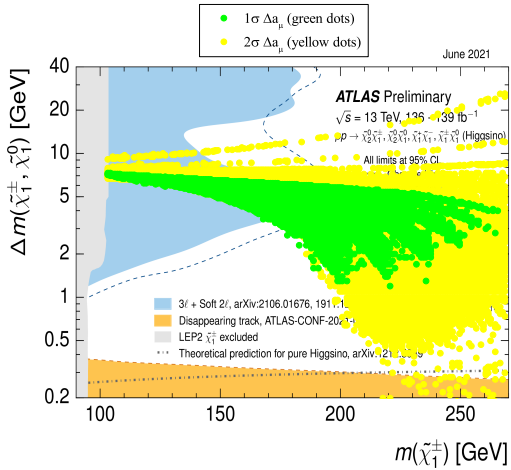}
        \caption{Illustration of the \fsu5 D-brane model points that can explain the BNL+FNAL 4.2$\sigma$ discrepancy on $\Delta a_{\mu}$ evaluated against the ATLAS constraints on electroweakinos for those spectra with higgsino LSP. The points are superimposed upon the ATLAS exclusion curves of Ref.~\cite{Aad:2019qnd} that addresses electroweakino production in compressed spectra, namely, higgsino. The $SU(5)$ breaking effects drive the wino to small values, engendering some tension with the lower LHC bound on chargino masses. However, as this Figure exhibits, there remains a large region that handily satisfies this constraint.}
        \label{fig:lhc}
\end{figure}

\begin{figure}[htp]
       \centering
        \includegraphics[width=0.5\textwidth]{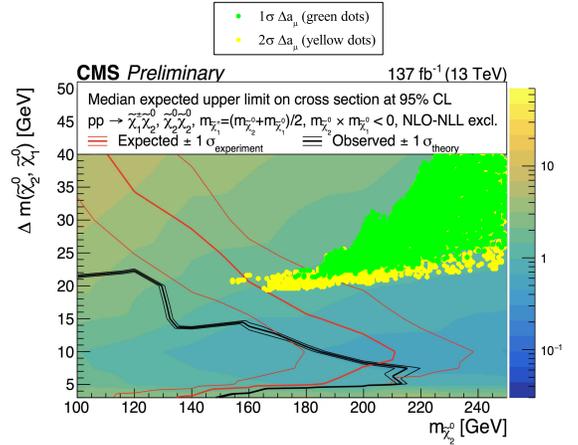}
        \caption{Depiction of the \fsu5 D-brane model points that can explain the BNL+FNAL 4.2$\sigma$ discrepancy on $\Delta a_{\mu}$ evaluated against the CMS constraints on electroweakinos for those spectra with higgsino LSP. The points are superimposed upon the CMS exclusion curves of Ref.~\cite{CMS:ew} that addresses electroweakino production in compressed spectra, namely, higgsino. In this plot for clarity we only show those points that survive the ATLAS electroweakino constraint of Ref.~\cite{Aad:2019qnd}, as displayed in FIG.~\ref{fig:lhc}. The CMS simplified model scenario applied is not an exact match for the \fsu5 D-brane model space, but it is nonetheless instructive as to how our model measures against the latest chargino constraints.}
        \label{fig:cms}
\end{figure}

\begin{figure*}[htp]
       \centering
        \includegraphics[width=1.0\textwidth]{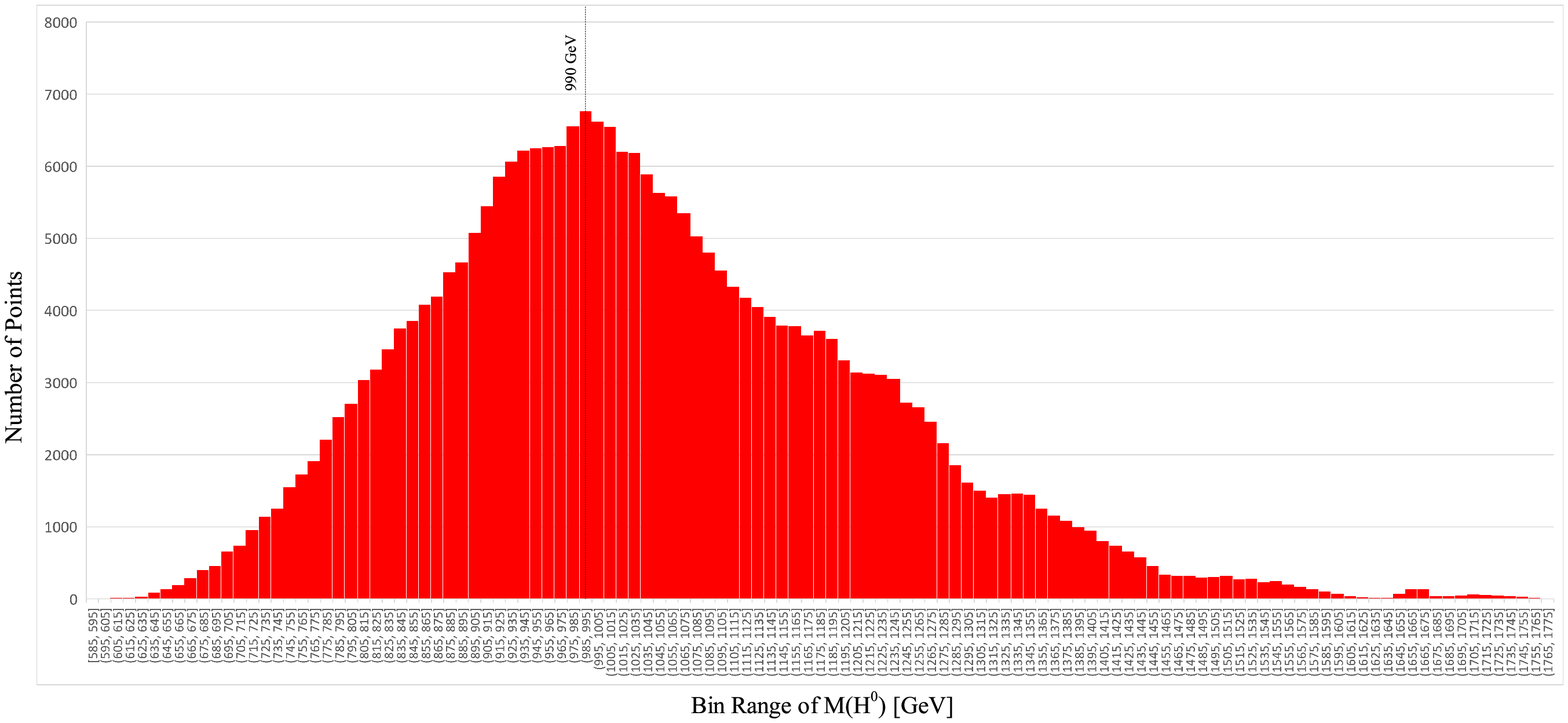}
        \caption{Histogram binned by counts of the heavy SUSY Higgs mass $M(H^0)$. All 261,562 points that satisfy the LHC electroweakino constraints and reside within either the BNL+FNAL 1$\sigma$ or $2\sigma$ limits on $\Delta a_{\mu}$ are counted in the bins. The bin width is 10~GeV. The peak lies in the $\left( 985,995 \right]$ bin, or $M(H^0) \simeq 1$~TeV.}
        \label{fig:histogram}
\end{figure*}

The computational results are constrained by filtering the data through current limits on pertinent beyond the Standard Model (BSM) experiments. A tolerance of $2\sigma$ is allowed around the branching ratio Br$(b \to s \gamma)$ of the b-quark decay~\cite{HeavyFlavorAveragingGroup:2012zzm} and branching ratio Br$(B_S^0 \to \mu^+ \mu^-)$ of the rare B-meson decay to a dimuon~\cite{:2012ct}. The WMAP 9-year~\cite{Hinshaw:2012aka} and 2015-18 Planck~\cite{Ade:2015xua,Aghanim:2018eyx} upper limit of about $\Omega h^2 \simeq 0.12$ is required, though no lower limit on the relic density is applied. A lower bound on the proton decay rate $\tau_p (p \to e^+ \pi^0) \ge 1.7 \times 10^{34}$~yrs~\cite{Takhistov:2016eqm} is also enforced. In summary, the points are filtered to ensure consistency with the following requirements:
\begin{eqnarray}
\nonumber
&&2.99 \le {\rm Br}(b \to s \gamma) \le 3.87 \times 10^{-4} \\ \nonumber
&&0.8 \le {\rm Br}(B_S^0 \to \mu^+ \mu^-) \le 6.2 \times 10^{-9} \\ \nonumber
&&0 \le \Omega h^2 \le 0.12 \\ \nonumber
&&\tau_p (p \to e^+ \pi^0) \ge 1.7 \times 10^{34}~{\rm yrs} \\ \nonumber
&&123 \le m_h \le 127~{\rm GeV} \\ \nonumber
\label{constraints}
\end{eqnarray}

A 2$\sigma$ experimental uncertainty and reasonable theoretical uncertainty is permitted around the observed light Higgs boson mass of $m_h = 125.09$~GeV~\cite{Aad:2012tfa, Chatrchyan:2012xdj}, providing a range of $123 \le m_h \le 127$~GeV on the computed total $m_h$. The total light Higgs boson mass calculation includes the vector-like particle contribution, which couples through the vector-like multiplet Yukawa  coupling. We assume a maximum coupling, implying the $(XD, XD^c)$ Yukawa coupling is $Y_{XD} = 0$ and the $(XU,XU^c)$ Yukawa coupling is $Y_{XU} = 1$, and furthermore, the trilinear coupling $A$-term is $A_{XD} = 0$ while the $(XU, XU^c)$ $A$-term is $A_{XU} = A_U$~\cite{Huo:2011zt,Li:2011ab}. These numerical values ensure a maximal coupling between the vector-like particles and light Higgs boson.

Given only an upper limit established on the abundance of the lightest neutralino $\widetilde{\chi}_1^0$, multi-component dark matter is generally necessary. Therefore, we rescale the spin-independent cross-section $\sigma_{SI}$ on nucleon-neutralino collisions as such:

\begin{eqnarray}
\sigma_{SI}^{\rm rescaled} = \sigma_{SI} \frac{\Omega h^2}{0.12}
\label{rescale}
\end{eqnarray}

The $SU(5)$ breaking effects drive the wino to near degeneracy with the lightest neutralino, generating higgsino LSPs. Compressed spectra with $M(\widetilde{\chi}_1^{\pm}) - M(\widetilde{\chi}_1^0) = 3 - 7$~GeV and $M(\widetilde{\chi}_2^0) < 0$ are typical conditions identifying higgsino spectra, and indeed these are the characteristics of the \fsu5 D-brane points. Twelve sample benchmark points are presented in TABLE~\ref{points}, with the higgsino LSP composition percent shown. Given the small wino, the most stringent LHC constraints on the \fsu5 D-brane model are ATLAS electroweakino production searches for higgsino LSP~\cite{Aad:2019qnd}. The small wino though does provide ample contribution to $\Delta a_{\mu}$(SUSY) to explain the BNL+FNAL 4.2$\sigma$ discrepancy. The D-brane model has a large region within both the 1$\sigma$ and $2\sigma$ limits around the BNL+FNAL $25.1 \times 10^{-10}$ central value. This is graphically illustrated in FIG.~\ref{fig:gluino}, where $\Delta a_{\mu}$(SUSY) is plot versus the gluino mass. The points are distinguished by satisfaction of the ATLAS light chargino constraint for those spectra with higgsino LSP~\cite{Aad:2019qnd}. The uncertainty on all $\Delta a_{\mu}$ calculations is about $\pm 2.4 \times 10^{-10}$. For the amount of $M_5$ and $M_{1X}$ mixing in Eq.~\ref{breaking}, $\Delta a_{\mu}$(SUSY) remains in the BNL+FNAL 1$\sigma$ range up to $M(\widetilde{g}) \approx 2.56$~TeV. It is clear in FIG.~\ref{fig:gluino} that the current LHC Run 2 gluino constraint of $M(\widetilde{g}) \gtrsim 2.25$~TeV~\cite{ATLAS:2018yhd,ATLAS:2020xgt,ATLAS:2017tmw,Sirunyan:2019xwh,Sirunyan:2019ctn,Sirunyan:2021mrs,CMS:2019tlp} correlates to the current chargino constraint for higgsino LSP since the BNL+FNAL 1$\sigma$ and 2$\sigma$ points that satisfy the ATLAS chargino constraint show a coincident lower bound of $M(\widetilde{g}) \simeq 2.25$~TeV.

To more accurately assess any tension with LHC electroweakino production constraints so that we can provide a frame of reference for our relatively small wino, we superimpose the D-brane model points of FIG.~\ref{fig:gluino} onto the ATLAS electroweakino production summary plot for higgsino LSP~\cite{Aad:2019qnd}, as displayed in FIG.~\ref{fig:lhc}, and also the CMS electroweakino production plot for higgsino LSP~\cite{CMS:ew}, as shown in FIG.~\ref{fig:cms}. The points are likewise separated into those that reside within either the BNL+FNAL 1$\sigma$ or 2$\sigma$ limits on $\Delta a_{\mu}$. Given the compressed nature of the D-brane spectra, these higgsino simplified models appear to be the only search regions capable of probing the D-brane model space that resolves the BNL+FNAL 4.2$\sigma$ discrepancy. In the CMS plot in FIG.~\ref{fig:cms}, for clarity we show only those points that satisfy the ATLAS chargino constraint of FIG.~\ref{fig:lhc}. The CMS simplified model scenario applied in FIG.~\ref{fig:cms} is not a precise fit for the \fsu5 D-brane model space, but it does give a good summary as to how our model stands against the current chargino constraints. The CMS analysis of Ref.~\cite{CMS:ew} also studies the higgsino model in the phenomenological MSSM (pMSSM), though constraints are given in terms of the wino mass as a function of $\mu$ and all points in our model are well beyond these exclusion limits due to the large $\mu$ term.

Other than the small wino, the only other class of measurements the D-brane model would seem to be experiencing tension with are the direct-detection spin-independent cross-sections, which we rescale for small $\Omega h^2$. However, given the difficulty with which higgsino LSPs can be detected, we are not alarmed by the larger $\sigma_{SI}^{\rm rescaled} \sim 10^{-9}$~pb cross-section.

Recently, ATLAS and CMS have completed searches for heavy resonances that would include heavy Higgs bosons found in an extended Higgs sector. The ATLAS search~\cite{ATLAS_heavy} involves b-jet and $\tau$-lepton final states, whereas the CMS search~\cite{CMS_heavy} focuses on $b$-quark pairs. All 261,562 points that pass the LHC chargino constraints on higgsino spectra and concurrently reside within either the BNL+FNAL 1$\sigma$ or $2\sigma$ limits on $\Delta a_{\mu}$ are binned in the histogram of FIG.~\ref{fig:histogram}, using 10~GeV bin widths. The histogram peak resides in the $\left( 985,995 \right]$~GeV bin, or $M(H^0) \simeq 1$~TeV.

\section{Conclusions}

The recent FNAL confirmation of the BNL discrepancy between the measured value of the anomalous magnetic moment of the muon $(g - 2)_{\mu}$ and the Standard Model (SM) prediction presents a combined deviation of 4.2$\sigma$. Given the possible confirmation of a statistically significant 5$\sigma$ discovery in forthcoming years, all natural GUT model candidates should be capable of elegantly explaining these experimental anomalies. Supersymmetry (SUSY) persists as one promising extension of the SM, hence we study in this work a merging of a realistic D-brane model with the supersymmetric GUT model flipped $SU(5)$ with extra TeV-scale string derived vector-like multiplets, referred to collectively as the \fsu5 D-brane model. Flipped $SU(5)$ has been shown to be both phenomenologically and cosmologically favorable, therefore, it is an ideal candidate to pursue whether it can indeed resolve the BNL+FNAL observed disparity with the SM calculations.

The supersymmetric GUT model \fsu5 produces a distinctive two-stage unification process. The primary unification occurs at the $SU(5) \times U(1)_X$ scale where $M_1 = M_2 = M_3$, then a secondary unification at the $SU(3)_C \times SU(2)_L$ scale. However, the large wino mass $M_2$ at $SU(5) \times U(1)_X$ due to the primary unification presents difficulties since a light wino at low-energy provides a large contribution to the muon $(g - 2)_{\mu}$. This dilemma can be resolved when the three chiral multiplets in the $SU(5)$ adjoint representation acquire vevs around the scale $SU(3)_C \times SU(2)_L$, and then the gaugino masses for $SU(3)_C \times SU(2)_L \times U(1)_{Y'}$ of $SU(5)$ can be split due to high-dimensional operators, leading to independent bino, wino, and gluino masses. These effects from $SU(5)$ breaking can drive the wino $M_2$ to small values at the electroweak scale, which in consequence generates a large contribution to the muon $(g - 2)_{\mu}$.

A deep analysis of the parameter space uncovered a region in the model that can explain the BNL+FNAL muon $(g - 2)_{\mu}$ measurements of  $\Delta a_{\mu} = a_{\mu} ({\rm Exp}) - a_{\mu} ({\rm SM}) = 25.1 \pm 5.9 \times 10^{-10}$. Calculations show that the model can produce an anomalous magnetic moment as large as $\Delta a_{\mu}({\rm SUSY}) = 22.3 \times 10^{-10}$ for a gluino mass of $M(\widetilde{g}) = 2281$~GeV, well within the 1$\sigma$ uncertainty on the observed value. Moreover, we completed computations for gluino masses as large as $M(\widetilde{g}) = 2560$~GeV, sufficient to ensure probing of the model space for several years to come at the LHC Run 2. We found that $\Delta a_{\mu}({\rm SUSY})$ can remain tucked just above the lower 1$\sigma$ bound of $\Delta a_{\mu}({\rm SUSY}) \simeq 19.0 \times 10^{-10}$ all the way up to the 2560~GeV gluino mass. The small $M_2$ does drive the chargino mass $M(\widetilde{\chi}_1^{\pm})$ to near degeneracy with the LSP mass $M(\widetilde{\chi}_1^0)$ along with a small negative neutralino mass $M(\widetilde{\chi}_2^0)$, generating higgsino dark matter. In addition to the favorable  $\Delta a_{\mu}({\rm SUSY})$, these same points in the model space are consistent with the observed light Higgs boson mass and other rare decay branching ratios, and also all LHC constraints, particularly those on electroweakinos, given the small chargino. Lastly, in light of recent LHC searches for heavy Higgs bosons in an extended Higgs sector, for this same region that can explain the BNL+FNAL measurements, we examined the heavy SUSY Higgs by plotting all viable $(g - 2)_{\mu}$ points in a histogram with 10~GeV bin widths and discovered that the histogram peak resides at $M(H^0) \simeq 1$~TeV.


\section{Acknowledgments}

Portions of this research were conducted with high performance computational resources provided 
by the Louisiana Optical Network Infrastructure (http://www.loni.org). This research was supported 
in part by the Projects 11475238, 11647601, and 11875062 supported 
by the National Natural Science Foundation of China (TL), 
by the Key Research Program of Frontier Science, Chinese Academy of Sciences (TL),
and by the DOE grant DE-FG02-13ER42020 (DVN). 


\bibliography{bibliography}

\begin{thebibliography}{106}
\expandafter\ifx\csname natexlab\endcsname\relax\def\natexlab#1{#1}\fi
\expandafter\ifx\csname bibnamefont\endcsname\relax
  \def\bibnamefont#1{#1}\fi
\expandafter\ifx\csname bibfnamefont\endcsname\relax
  \def\bibfnamefont#1{#1}\fi
\expandafter\ifx\csname citenamefont\endcsname\relax
  \def\citenamefont#1{#1}\fi
\expandafter\ifx\csname url\endcsname\relax
  \def\url#1{\texttt{#1}}\fi
\expandafter\ifx\csname urlprefix\endcsname\relax\def\urlprefix{URL }\fi
\providecommand{\bibinfo}[2]{#2}
\providecommand{\eprint}[2][]{\url{#2}}

\bibitem[{\citenamefont{Bennett et~al.}(2006)}]{bennett:2006fi}
\bibinfo{author}{\bibfnamefont{G.~W.} \bibnamefont{Bennett}}
  \bibnamefont{et~al.} (\bibinfo{collaboration}{Muon $g-2$}),
  {``}\bibinfo{title}{{Final Report of the Muon E821 Anomalous Magnetic Moment
  Measurement at BNL}},{''} \bibinfo{journal}{Phys. Rev.}
  \textbf{\bibinfo{volume}{D73}}, \bibinfo{pages}{072003}
  (\bibinfo{year}{2006}), \eprint{hep-ex/0602035}.

\bibitem[{\citenamefont{Davier et~al.}(2011)\citenamefont{Davier, Hoecker,
  Malaescu, and Zhang}}]{Davier:2010nc}
\bibinfo{author}{\bibfnamefont{M.}~\bibnamefont{Davier}},
  \bibinfo{author}{\bibfnamefont{A.}~\bibnamefont{Hoecker}},
  \bibinfo{author}{\bibfnamefont{B.}~\bibnamefont{Malaescu}}, \bibnamefont{and}
  \bibinfo{author}{\bibfnamefont{Z.}~\bibnamefont{Zhang}},
  {``}\bibinfo{title}{{Reevaluation of the Hadronic Contributions to the Muon
  g-2 and to alpha(MZ)}},{''} \bibinfo{journal}{Eur.Phys.J.}
  \textbf{\bibinfo{volume}{C71}}, \bibinfo{pages}{1515} (\bibinfo{year}{2011}),
  \eprint{1010.4180}.

\bibitem[{\citenamefont{Davier et~al.}(2017)\citenamefont{Davier, Hoecker,
  Malaescu, and Zhang}}]{davier:2017zfy}
\bibinfo{author}{\bibfnamefont{M.}~\bibnamefont{Davier}},
  \bibinfo{author}{\bibfnamefont{A.}~\bibnamefont{Hoecker}},
  \bibinfo{author}{\bibfnamefont{B.}~\bibnamefont{Malaescu}}, \bibnamefont{and}
  \bibinfo{author}{\bibfnamefont{Z.}~\bibnamefont{Zhang}},
  {``}\bibinfo{title}{{Reevaluation of the hadronic vacuum polarisation
  contributions to the Standard Model predictions of the muon $g-2$ and
  ${\alpha (m_Z^2)}$ using newest hadronic cross-section data}},{''}
  \bibinfo{journal}{Eur. Phys. J.} \textbf{\bibinfo{volume}{C77}},
  \bibinfo{pages}{827} (\bibinfo{year}{2017}), \eprint{1706.09436}.

\bibitem[{\citenamefont{Keshavarzi et~al.}(2018)\citenamefont{Keshavarzi,
  Nomura, and Teubner}}]{keshavarzi:2018mgv}
\bibinfo{author}{\bibfnamefont{A.}~\bibnamefont{Keshavarzi}},
  \bibinfo{author}{\bibfnamefont{D.}~\bibnamefont{Nomura}}, \bibnamefont{and}
  \bibinfo{author}{\bibfnamefont{T.}~\bibnamefont{Teubner}},
  {``}\bibinfo{title}{{Muon $g-2$ and $\alpha(M_Z^2)$: a new data-based
  analysis}},{''} \bibinfo{journal}{Phys. Rev.} \textbf{\bibinfo{volume}{D97}},
  \bibinfo{pages}{114025} (\bibinfo{year}{2018}), \eprint{1802.02995}.

\bibitem[{\citenamefont{Colangelo et~al.}(2019)\citenamefont{Colangelo,
  Hoferichter, and Stoffer}}]{colangelo:2018mtw}
\bibinfo{author}{\bibfnamefont{G.}~\bibnamefont{Colangelo}},
  \bibinfo{author}{\bibfnamefont{M.}~\bibnamefont{Hoferichter}},
  \bibnamefont{and} \bibinfo{author}{\bibfnamefont{P.}~\bibnamefont{Stoffer}},
  {``}\bibinfo{title}{{Two-pion contribution to hadronic vacuum
  polarization}},{''} \bibinfo{journal}{JHEP} \textbf{\bibinfo{volume}{02}},
  \bibinfo{pages}{006} (\bibinfo{year}{2019}), \eprint{1810.00007}.

\bibitem[{\citenamefont{Hoferichter et~al.}(2019)\citenamefont{Hoferichter,
  Hoid, and Kubis}}]{hoferichter:2019gzf}
\bibinfo{author}{\bibfnamefont{M.}~\bibnamefont{Hoferichter}},
  \bibinfo{author}{\bibfnamefont{B.-L.} \bibnamefont{Hoid}}, \bibnamefont{and}
  \bibinfo{author}{\bibfnamefont{B.}~\bibnamefont{Kubis}},
  {``}\bibinfo{title}{{Three-pion contribution to hadronic vacuum
  polarization}},{''} \bibinfo{journal}{JHEP} \textbf{\bibinfo{volume}{08}},
  \bibinfo{pages}{137} (\bibinfo{year}{2019}), \eprint{1907.01556}.

\bibitem[{\citenamefont{Davier et~al.}(2020)\citenamefont{Davier, Hoecker,
  Malaescu, and Zhang}}]{davier:2019can}
\bibinfo{author}{\bibfnamefont{M.}~\bibnamefont{Davier}},
  \bibinfo{author}{\bibfnamefont{A.}~\bibnamefont{Hoecker}},
  \bibinfo{author}{\bibfnamefont{B.}~\bibnamefont{Malaescu}}, \bibnamefont{and}
  \bibinfo{author}{\bibfnamefont{Z.}~\bibnamefont{Zhang}},
  {``}\bibinfo{title}{{A new evaluation of the hadronic vacuum polarisation
  contributions to the muon anomalous magnetic moment and to
  $\mathbf{\boldsymbol\alpha(m_Z^2)}$}},{''} \bibinfo{journal}{Eur. Phys. J.}
  \textbf{\bibinfo{volume}{C80}}, \bibinfo{pages}{241} (\bibinfo{year}{2020}),
  \bibinfo{note}{[Erratum: Eur. Phys. J. {\bf C80}, 410 (2020)]},
  \eprint{1908.00921}.

\bibitem[{\citenamefont{Keshavarzi et~al.}(2020)\citenamefont{Keshavarzi,
  Nomura, and Teubner}}]{keshavarzi:2019abf}
\bibinfo{author}{\bibfnamefont{A.}~\bibnamefont{Keshavarzi}},
  \bibinfo{author}{\bibfnamefont{D.}~\bibnamefont{Nomura}}, \bibnamefont{and}
  \bibinfo{author}{\bibfnamefont{T.}~\bibnamefont{Teubner}},
  {``}\bibinfo{title}{{The $g-2$ of charged leptons, $\alpha(M_Z^2)$ and the
  hyperfine splitting of muonium}},{''} \bibinfo{journal}{Phys. Rev.}
  \textbf{\bibinfo{volume}{D101}}, \bibinfo{pages}{014029}
  (\bibinfo{year}{2020}), \eprint{1911.00367}.

\bibitem[{\citenamefont{Kurz et~al.}(2014)\citenamefont{Kurz, Liu, Marquard,
  and Steinhauser}}]{kurz:2014wya}
\bibinfo{author}{\bibfnamefont{A.}~\bibnamefont{Kurz}},
  \bibinfo{author}{\bibfnamefont{T.}~\bibnamefont{Liu}},
  \bibinfo{author}{\bibfnamefont{P.}~\bibnamefont{Marquard}}, \bibnamefont{and}
  \bibinfo{author}{\bibfnamefont{M.}~\bibnamefont{Steinhauser}},
  {``}\bibinfo{title}{{Hadronic contribution to the muon anomalous magnetic
  moment to next-to-next-to-leading order}},{''} \bibinfo{journal}{Phys. Lett.}
  \textbf{\bibinfo{volume}{B734}}, \bibinfo{pages}{144} (\bibinfo{year}{2014}),
  \eprint{1403.6400}.

\bibitem[{\citenamefont{G\'erardin et~al.}(2019)\citenamefont{G\'erardin, C\`e,
  von Hippel, H{\"o}rz, Meyer, Mohler, Ottnad, Wilhelm, and
  Wittig}}]{gerardin:2019rua}
\bibinfo{author}{\bibfnamefont{A.}~\bibnamefont{G\'erardin}},
  \bibinfo{author}{\bibfnamefont{M.}~\bibnamefont{C\`e}},
  \bibinfo{author}{\bibfnamefont{G.}~\bibnamefont{von Hippel}},
  \bibinfo{author}{\bibfnamefont{B.}~\bibnamefont{H{\"o}rz}},
  \bibinfo{author}{\bibfnamefont{H.~B.} \bibnamefont{Meyer}},
  \bibinfo{author}{\bibfnamefont{D.}~\bibnamefont{Mohler}},
  \bibinfo{author}{\bibfnamefont{K.}~\bibnamefont{Ottnad}},
  \bibinfo{author}{\bibfnamefont{J.}~\bibnamefont{Wilhelm}}, \bibnamefont{and}
  \bibinfo{author}{\bibfnamefont{H.}~\bibnamefont{Wittig}},
  {``}\bibinfo{title}{{The leading hadronic contribution to $(g-2)_\mu$ from
  lattice QCD with $N_{\rm f}=2+1$ flavours of O($a$) improved Wilson
  quarks}},{''} \bibinfo{journal}{Phys. Rev.} \textbf{\bibinfo{volume}{D100}},
  \bibinfo{pages}{014510} (\bibinfo{year}{2019}), \eprint{1904.03120}.

\bibitem[{\citenamefont{Aubin et~al.}(2020)\citenamefont{Aubin, Blum, Tu,
  Golterman, Jung, and Peris}}]{Aubin:2019usy}
\bibinfo{author}{\bibfnamefont{C.}~\bibnamefont{Aubin}},
  \bibinfo{author}{\bibfnamefont{T.}~\bibnamefont{Blum}},
  \bibinfo{author}{\bibfnamefont{C.}~\bibnamefont{Tu}},
  \bibinfo{author}{\bibfnamefont{M.}~\bibnamefont{Golterman}},
  \bibinfo{author}{\bibfnamefont{C.}~\bibnamefont{Jung}}, \bibnamefont{and}
  \bibinfo{author}{\bibfnamefont{S.}~\bibnamefont{Peris}},
  {``}\bibinfo{title}{{Light quark vacuum polarization at the physical point
  and contribution to the muon $g-2$}},{''} \bibinfo{journal}{Phys. Rev.}
  \textbf{\bibinfo{volume}{D101}}, \bibinfo{pages}{014503}
  (\bibinfo{year}{2020}), \eprint{1905.09307}.

\bibitem[{\citenamefont{Giusti and Simula}(2019)}]{giusti:2019hkz}
\bibinfo{author}{\bibfnamefont{D.}~\bibnamefont{Giusti}} \bibnamefont{and}
  \bibinfo{author}{\bibfnamefont{S.}~\bibnamefont{Simula}},
  {``}\bibinfo{title}{{Lepton anomalous magnetic moments in Lattice
  QCD+QED}},{''} \bibinfo{journal}{PoS} \textbf{\bibinfo{volume}{LATTICE2019}},
  \bibinfo{pages}{104} (\bibinfo{year}{2019}), \eprint{1910.03874}.

\bibitem[{\citenamefont{Melnikov and Vainshtein}(2004)}]{melnikov:2003xd}
\bibinfo{author}{\bibfnamefont{K.}~\bibnamefont{Melnikov}} \bibnamefont{and}
  \bibinfo{author}{\bibfnamefont{A.}~\bibnamefont{Vainshtein}},
  {``}\bibinfo{title}{{Hadronic light-by-light scattering contribution to the
  muon anomalous magnetic moment revisited}},{''} \bibinfo{journal}{Phys. Rev.}
  \textbf{\bibinfo{volume}{D70}}, \bibinfo{pages}{113006}
  (\bibinfo{year}{2004}), \eprint{hep-ph/0312226}.

\bibitem[{\citenamefont{Masjuan and
  S{\'a}nchez-Puertas}(2017)}]{masjuan:2017tvw}
\bibinfo{author}{\bibfnamefont{P.}~\bibnamefont{Masjuan}} \bibnamefont{and}
  \bibinfo{author}{\bibfnamefont{P.}~\bibnamefont{S{\'a}nchez-Puertas}},
  {``}\bibinfo{title}{{Pseudoscalar-pole contribution to the $(g_{\mu}-2)$: a
  rational approach}},{''} \bibinfo{journal}{Phys. Rev.}
  \textbf{\bibinfo{volume}{D95}}, \bibinfo{pages}{054026}
  (\bibinfo{year}{2017}), \eprint{1701.05829}.

\bibitem[{\citenamefont{Colangelo et~al.}(2017)\citenamefont{Colangelo,
  Hoferichter, Procura, and Stoffer}}]{Colangelo:2017fiz}
\bibinfo{author}{\bibfnamefont{G.}~\bibnamefont{Colangelo}},
  \bibinfo{author}{\bibfnamefont{M.}~\bibnamefont{Hoferichter}},
  \bibinfo{author}{\bibfnamefont{M.}~\bibnamefont{Procura}}, \bibnamefont{and}
  \bibinfo{author}{\bibfnamefont{P.}~\bibnamefont{Stoffer}},
  {``}\bibinfo{title}{{Dispersion relation for hadronic light-by-light
  scattering: two-pion contributions}},{''} \bibinfo{journal}{JHEP}
  \textbf{\bibinfo{volume}{04}}, \bibinfo{pages}{161} (\bibinfo{year}{2017}),
  \eprint{1702.07347}.

\bibitem[{\citenamefont{Hoferichter et~al.}(2018)\citenamefont{Hoferichter,
  Hoid, Kubis, Leupold, and Schneider}}]{hoferichter:2018kwz}
\bibinfo{author}{\bibfnamefont{M.}~\bibnamefont{Hoferichter}},
  \bibinfo{author}{\bibfnamefont{B.-L.} \bibnamefont{Hoid}},
  \bibinfo{author}{\bibfnamefont{B.}~\bibnamefont{Kubis}},
  \bibinfo{author}{\bibfnamefont{S.}~\bibnamefont{Leupold}}, \bibnamefont{and}
  \bibinfo{author}{\bibfnamefont{S.~P.} \bibnamefont{Schneider}},
  {``}\bibinfo{title}{{Dispersion relation for hadronic light-by-light
  scattering: pion pole}},{''} \bibinfo{journal}{JHEP}
  \textbf{\bibinfo{volume}{10}}, \bibinfo{pages}{141} (\bibinfo{year}{2018}),
  \eprint{1808.04823}.

\bibitem[{\citenamefont{G{\'e}rardin et~al.}(2019)\citenamefont{G{\'e}rardin,
  Meyer, and Nyffeler}}]{gerardin:2019vio}
\bibinfo{author}{\bibfnamefont{A.}~\bibnamefont{G{\'e}rardin}},
  \bibinfo{author}{\bibfnamefont{H.~B.} \bibnamefont{Meyer}}, \bibnamefont{and}
  \bibinfo{author}{\bibfnamefont{A.}~\bibnamefont{Nyffeler}},
  {``}\bibinfo{title}{{Lattice calculation of the pion transition form factor
  with $N_f=2+1$ Wilson quarks}},{''} \bibinfo{journal}{Phys. Rev.}
  \textbf{\bibinfo{volume}{D100}}, \bibinfo{pages}{034520}
  (\bibinfo{year}{2019}), \eprint{1903.09471}.

\bibitem[{\citenamefont{Bijnens et~al.}(2019)\citenamefont{Bijnens,
  Hermansson-Truedsson, and Rodr{\'i}guez-S{\'a}nchez}}]{bijnens:2019ghy}
\bibinfo{author}{\bibfnamefont{J.}~\bibnamefont{Bijnens}},
  \bibinfo{author}{\bibfnamefont{N.}~\bibnamefont{Hermansson-Truedsson}},
  \bibnamefont{and}
  \bibinfo{author}{\bibfnamefont{A.}~\bibnamefont{Rodr{\'i}guez-S{\'a}nchez}},
  {``}\bibinfo{title}{{Short-distance constraints for the HLbL contribution to
  the muon anomalous magnetic moment}},{''} \bibinfo{journal}{Phys. Lett.}
  \textbf{\bibinfo{volume}{B798}}, \bibinfo{pages}{134994}
  (\bibinfo{year}{2019}), \eprint{1908.03331}.

\bibitem[{\citenamefont{Colangelo et~al.}(2020)\citenamefont{Colangelo,
  Hagelstein, Hoferichter, Laub, and Stoffer}}]{colangelo:2019uex}
\bibinfo{author}{\bibfnamefont{G.}~\bibnamefont{Colangelo}},
  \bibinfo{author}{\bibfnamefont{F.}~\bibnamefont{Hagelstein}},
  \bibinfo{author}{\bibfnamefont{M.}~\bibnamefont{Hoferichter}},
  \bibinfo{author}{\bibfnamefont{L.}~\bibnamefont{Laub}}, \bibnamefont{and}
  \bibinfo{author}{\bibfnamefont{P.}~\bibnamefont{Stoffer}},
  {``}\bibinfo{title}{{Longitudinal short-distance constraints for the hadronic
  light-by-light contribution to $(g-2)_\mu$ with large-$N_c$ Regge
  models}},{''} \bibinfo{journal}{JHEP} \textbf{\bibinfo{volume}{03}},
  \bibinfo{pages}{101} (\bibinfo{year}{2020}), \eprint{1910.13432}.

\bibitem[{\citenamefont{Pauk and Vanderhaeghen}(2014)}]{pauk:2014rta}
\bibinfo{author}{\bibfnamefont{V.}~\bibnamefont{Pauk}} \bibnamefont{and}
  \bibinfo{author}{\bibfnamefont{M.}~\bibnamefont{Vanderhaeghen}},
  {``}\bibinfo{title}{{Single meson contributions to the muon`s anomalous
  magnetic moment}},{''} \bibinfo{journal}{Eur. Phys. J.}
  \textbf{\bibinfo{volume}{C74}}, \bibinfo{pages}{3008} (\bibinfo{year}{2014}),
  \eprint{1401.0832}.

\bibitem[{\citenamefont{Danilkin and Vanderhaeghen}(2017)}]{danilkin:2016hnh}
\bibinfo{author}{\bibfnamefont{I.}~\bibnamefont{Danilkin}} \bibnamefont{and}
  \bibinfo{author}{\bibfnamefont{M.}~\bibnamefont{Vanderhaeghen}},
  {``}\bibinfo{title}{{Light-by-light scattering sum rules in light of new
  data}},{''} \bibinfo{journal}{Phys. Rev.} \textbf{\bibinfo{volume}{D95}},
  \bibinfo{pages}{014019} (\bibinfo{year}{2017}), \eprint{1611.04646}.

\bibitem[{\citenamefont{Jegerlehner}(2017)}]{jegerlehner:2017gek}
\bibinfo{author}{\bibfnamefont{F.}~\bibnamefont{Jegerlehner}},
  {``}\bibinfo{title}{{The Anomalous Magnetic Moment of the Muon}},{''}
  \bibinfo{journal}{Springer Tracts Mod. Phys.} \textbf{\bibinfo{volume}{274}},
  \bibinfo{pages}{1} (\bibinfo{year}{2017}).

\bibitem[{\citenamefont{Knecht et~al.}(2018)\citenamefont{Knecht, Narison,
  Rabemananjara, and Rabetiarivony}}]{knecht:2018sci}
\bibinfo{author}{\bibfnamefont{M.}~\bibnamefont{Knecht}},
  \bibinfo{author}{\bibfnamefont{S.}~\bibnamefont{Narison}},
  \bibinfo{author}{\bibfnamefont{A.}~\bibnamefont{Rabemananjara}},
  \bibnamefont{and}
  \bibinfo{author}{\bibfnamefont{D.}~\bibnamefont{Rabetiarivony}},
  {``}\bibinfo{title}{{Scalar meson contributions to $a_\mu$ from hadronic
  light-by-light scattering}},{''} \bibinfo{journal}{Phys. Lett.}
  \textbf{\bibinfo{volume}{B787}}, \bibinfo{pages}{111} (\bibinfo{year}{2018}),
  \eprint{1808.03848}.

\bibitem[{\citenamefont{Eichmann et~al.}(2020)\citenamefont{Eichmann, Fischer,
  and Williams}}]{eichmann:2019bqf}
\bibinfo{author}{\bibfnamefont{G.}~\bibnamefont{Eichmann}},
  \bibinfo{author}{\bibfnamefont{C.~S.} \bibnamefont{Fischer}},
  \bibnamefont{and} \bibinfo{author}{\bibfnamefont{R.}~\bibnamefont{Williams}},
  {``}\bibinfo{title}{{Kaon-box contribution to the anomalous magnetic moment
  of the muon}},{''} \bibinfo{journal}{Phys. Rev.}
  \textbf{\bibinfo{volume}{D101}}, \bibinfo{pages}{054015}
  (\bibinfo{year}{2020}), \eprint{1910.06795}.

\bibitem[{\citenamefont{Roig and S{\'a}nchez-Puertas}(2020)}]{roig:2019reh}
\bibinfo{author}{\bibfnamefont{P.}~\bibnamefont{Roig}} \bibnamefont{and}
  \bibinfo{author}{\bibfnamefont{P.}~\bibnamefont{S{\'a}nchez-Puertas}},
  {``}\bibinfo{title}{{Axial-vector exchange contribution to the hadronic
  light-by-light piece of the muon anomalous magnetic moment}},{''}
  \bibinfo{journal}{Phys. Rev.} \textbf{\bibinfo{volume}{D101}},
  \bibinfo{pages}{074019} (\bibinfo{year}{2020}), \eprint{1910.02881}.

\bibitem[{\citenamefont{Blum et~al.}(2020)\citenamefont{Blum, Christ, Hayakawa,
  Izubuchi, Jin, Jung, and Lehner}}]{Blum:2019ugy}
\bibinfo{author}{\bibfnamefont{T.}~\bibnamefont{Blum}},
  \bibinfo{author}{\bibfnamefont{N.}~\bibnamefont{Christ}},
  \bibinfo{author}{\bibfnamefont{M.}~\bibnamefont{Hayakawa}},
  \bibinfo{author}{\bibfnamefont{T.}~\bibnamefont{Izubuchi}},
  \bibinfo{author}{\bibfnamefont{L.}~\bibnamefont{Jin}},
  \bibinfo{author}{\bibfnamefont{C.}~\bibnamefont{Jung}}, \bibnamefont{and}
  \bibinfo{author}{\bibfnamefont{C.}~\bibnamefont{Lehner}},
  {``}\bibinfo{title}{{The hadronic light-by-light scattering contribution to
  the muon anomalous magnetic moment from lattice QCD}},{''}
  \bibinfo{journal}{Phys. Rev. Lett.} \textbf{\bibinfo{volume}{124}},
  \bibinfo{pages}{132002} (\bibinfo{year}{2020}), \eprint{1911.08123}.

\bibitem[{\citenamefont{Colangelo et~al.}(2014)\citenamefont{Colangelo,
  Hoferichter, Nyffeler, Passera, and Stoffer}}]{colangelo:2014qya}
\bibinfo{author}{\bibfnamefont{G.}~\bibnamefont{Colangelo}},
  \bibinfo{author}{\bibfnamefont{M.}~\bibnamefont{Hoferichter}},
  \bibinfo{author}{\bibfnamefont{A.}~\bibnamefont{Nyffeler}},
  \bibinfo{author}{\bibfnamefont{M.}~\bibnamefont{Passera}}, \bibnamefont{and}
  \bibinfo{author}{\bibfnamefont{P.}~\bibnamefont{Stoffer}},
  {``}\bibinfo{title}{{Remarks on higher-order hadronic corrections to the muon
  $g-2$}},{''} \bibinfo{journal}{Phys. Lett.} \textbf{\bibinfo{volume}{B735}},
  \bibinfo{pages}{90} (\bibinfo{year}{2014}), \eprint{1403.7512}.

\bibitem[{\citenamefont{Aoyama et~al.}(2012)\citenamefont{Aoyama, Hayakawa,
  Kinoshita, and Nio}}]{Aoyama:2012wk}
\bibinfo{author}{\bibfnamefont{T.}~\bibnamefont{Aoyama}},
  \bibinfo{author}{\bibfnamefont{M.}~\bibnamefont{Hayakawa}},
  \bibinfo{author}{\bibfnamefont{T.}~\bibnamefont{Kinoshita}},
  \bibnamefont{and} \bibinfo{author}{\bibfnamefont{M.}~\bibnamefont{Nio}},
  {``}\bibinfo{title}{{Complete Tenth-Order QED Contribution to the Muon
  g-2}},{''} \bibinfo{journal}{Phys.Rev.Lett.} \textbf{\bibinfo{volume}{109}},
  \bibinfo{pages}{111808} (\bibinfo{year}{2012}), \eprint{1205.5370}.

\bibitem[{\citenamefont{Aoyama et~al.}(2019)\citenamefont{Aoyama, Kinoshita,
  and Nio}}]{Aoyama:2019ryr}
\bibinfo{author}{\bibfnamefont{T.}~\bibnamefont{Aoyama}},
  \bibinfo{author}{\bibfnamefont{T.}~\bibnamefont{Kinoshita}},
  \bibnamefont{and} \bibinfo{author}{\bibfnamefont{M.}~\bibnamefont{Nio}},
  {``}\bibinfo{title}{{Theory of the Anomalous Magnetic Moment of the
  Electron}},{''} \bibinfo{journal}{Atoms} \textbf{\bibinfo{volume}{7}},
  \bibinfo{pages}{28} (\bibinfo{year}{2019}).

\bibitem[{\citenamefont{Czarnecki et~al.}(2003)\citenamefont{Czarnecki,
  Marciano, and Vainshtein}}]{czarnecki:2002nt}
\bibinfo{author}{\bibfnamefont{A.}~\bibnamefont{Czarnecki}},
  \bibinfo{author}{\bibfnamefont{W.~J.} \bibnamefont{Marciano}},
  \bibnamefont{and}
  \bibinfo{author}{\bibfnamefont{A.}~\bibnamefont{Vainshtein}},
  {``}\bibinfo{title}{{Refinements in electroweak contributions to the muon
  anomalous magnetic moment}},{''} \bibinfo{journal}{Phys. Rev.}
  \textbf{\bibinfo{volume}{D67}}, \bibinfo{pages}{073006}
  (\bibinfo{year}{2003}), \bibinfo{note}{[Erratum: Phys. Rev. {\bf D73}, 119901
  (2006)]}, \eprint{hep-ph/0212229}.

\bibitem[{\citenamefont{Gnendiger et~al.}(2013)\citenamefont{Gnendiger,
  St{\"o}ckinger, and St{\"o}ckinger-Kim}}]{gnendiger:2013pva}
\bibinfo{author}{\bibfnamefont{C.}~\bibnamefont{Gnendiger}},
  \bibinfo{author}{\bibfnamefont{D.}~\bibnamefont{St{\"o}ckinger}},
  \bibnamefont{and}
  \bibinfo{author}{\bibfnamefont{H.}~\bibnamefont{St{\"o}ckinger-Kim}},
  {``}\bibinfo{title}{{The electroweak contributions to $(g-2)_\mu$ after the
  Higgs boson mass measurement}},{''} \bibinfo{journal}{Phys. Rev.}
  \textbf{\bibinfo{volume}{D88}}, \bibinfo{pages}{053005}
  (\bibinfo{year}{2013}), \eprint{1306.5546}.

\bibitem[{\citenamefont{Abi et~al.}(2021)}]{Abi:2021gix}
\bibinfo{author}{\bibfnamefont{B.}~\bibnamefont{Abi}} \bibnamefont{et~al.}
  (\bibinfo{collaboration}{Muon g-2}), {``}\bibinfo{title}{{Measurement of the
  Positive Muon Anomalous Magnetic Moment to 0.46~ppm}},{''}
  \bibinfo{journal}{Phys. Rev. Lett.} \textbf{\bibinfo{volume}{126}},
  \bibinfo{pages}{141801} (\bibinfo{year}{2021}), \eprint{2104.03281}.

\bibitem[{\citenamefont{Berkooz et~al.}(1996)\citenamefont{Berkooz, Douglas,
  and Leigh}}]{Berkooz:1996km}
\bibinfo{author}{\bibfnamefont{M.}~\bibnamefont{Berkooz}},
  \bibinfo{author}{\bibfnamefont{M.~R.} \bibnamefont{Douglas}},
  \bibnamefont{and} \bibinfo{author}{\bibfnamefont{R.~G.} \bibnamefont{Leigh}},
  {``}\bibinfo{title}{{Branes intersecting at angles}},{''}
  \bibinfo{journal}{Nucl.Phys.} \textbf{\bibinfo{volume}{B480}},
  \bibinfo{pages}{265} (\bibinfo{year}{1996}), \eprint{hep-th/9606139}.

\bibitem[{\citenamefont{Cvetic et~al.}(2001{\natexlab{a}})\citenamefont{Cvetic,
  Shiu, and Uranga}}]{Cvetic:2001tj}
\bibinfo{author}{\bibfnamefont{M.}~\bibnamefont{Cvetic}},
  \bibinfo{author}{\bibfnamefont{G.}~\bibnamefont{Shiu}}, \bibnamefont{and}
  \bibinfo{author}{\bibfnamefont{A.~M.} \bibnamefont{Uranga}},
  {``}\bibinfo{title}{{Three family supersymmetric standard - like models from
  intersecting brane worlds}},{''} \bibinfo{journal}{Phys.Rev.Lett.}
  \textbf{\bibinfo{volume}{87}}, \bibinfo{pages}{201801}
  (\bibinfo{year}{2001}{\natexlab{a}}), \eprint{hep-th/0107143}.

\bibitem[{\citenamefont{Cvetic et~al.}(2001{\natexlab{b}})\citenamefont{Cvetic,
  Shiu, and Uranga}}]{Cvetic:2001nr}
\bibinfo{author}{\bibfnamefont{M.}~\bibnamefont{Cvetic}},
  \bibinfo{author}{\bibfnamefont{G.}~\bibnamefont{Shiu}}, \bibnamefont{and}
  \bibinfo{author}{\bibfnamefont{A.~M.} \bibnamefont{Uranga}},
  {``}\bibinfo{title}{{Chiral four-dimensional N=1 supersymmetric type 2A
  orientifolds from intersecting D6 branes}},{''} \bibinfo{journal}{Nucl.Phys.}
  \textbf{\bibinfo{volume}{B615}}, \bibinfo{pages}{3}
  (\bibinfo{year}{2001}{\natexlab{b}}), \eprint{hep-th/0107166}.

\bibitem[{\citenamefont{Blumenhagen et~al.}(2001)\citenamefont{Blumenhagen,
  Kors, Lust, and Ott}}]{Blumenhagen:2001te}
\bibinfo{author}{\bibfnamefont{R.}~\bibnamefont{Blumenhagen}},
  \bibinfo{author}{\bibfnamefont{B.}~\bibnamefont{Kors}},
  \bibinfo{author}{\bibfnamefont{D.}~\bibnamefont{Lust}}, \bibnamefont{and}
  \bibinfo{author}{\bibfnamefont{T.}~\bibnamefont{Ott}},
  {``}\bibinfo{title}{{The standard model from stable intersecting brane world
  orbifolds}},{''} \bibinfo{journal}{Nucl.Phys.}
  \textbf{\bibinfo{volume}{B616}}, \bibinfo{pages}{3} (\bibinfo{year}{2001}),
  \eprint{hep-th/0107138}.

\bibitem[{\citenamefont{Ibanez et~al.}(2001)\citenamefont{Ibanez, Marchesano,
  and Rabadan}}]{Ibanez:2001nd}
\bibinfo{author}{\bibfnamefont{L.~E.} \bibnamefont{Ibanez}},
  \bibinfo{author}{\bibfnamefont{F.}~\bibnamefont{Marchesano}},
  \bibnamefont{and} \bibinfo{author}{\bibfnamefont{R.}~\bibnamefont{Rabadan}},
  {``}\bibinfo{title}{{Getting just the standard model at intersecting
  branes}},{''} \bibinfo{journal}{JHEP} \textbf{\bibinfo{volume}{0111}},
  \bibinfo{pages}{002} (\bibinfo{year}{2001}), \eprint{hep-th/0105155}.

\bibitem[{\citenamefont{Cvetic et~al.}(2003)\citenamefont{Cvetic,
  Papadimitriou, and Shiu}}]{Cvetic:2002pj}
\bibinfo{author}{\bibfnamefont{M.}~\bibnamefont{Cvetic}},
  \bibinfo{author}{\bibfnamefont{I.}~\bibnamefont{Papadimitriou}},
  \bibnamefont{and} \bibinfo{author}{\bibfnamefont{G.}~\bibnamefont{Shiu}},
  {``}\bibinfo{title}{{Supersymmetric three family SU(5) grand unified models
  from type IIA orientifolds with intersecting D6-branes}},{''}
  \bibinfo{journal}{Nucl.Phys.} \textbf{\bibinfo{volume}{B659}},
  \bibinfo{pages}{193} (\bibinfo{year}{2003}), \eprint{hep-th/0212177}.

\bibitem[{\citenamefont{Blumenhagen et~al.}(2005)\citenamefont{Blumenhagen,
  Cvetic, Langacker, and Shiu}}]{Blumenhagen:2005mu}
\bibinfo{author}{\bibfnamefont{R.}~\bibnamefont{Blumenhagen}},
  \bibinfo{author}{\bibfnamefont{M.}~\bibnamefont{Cvetic}},
  \bibinfo{author}{\bibfnamefont{P.}~\bibnamefont{Langacker}},
  \bibnamefont{and} \bibinfo{author}{\bibfnamefont{G.}~\bibnamefont{Shiu}},
  {``}\bibinfo{title}{{Toward realistic intersecting D-brane models}},{''}
  \bibinfo{journal}{Ann.Rev.Nucl.Part.Sci.} \textbf{\bibinfo{volume}{55}},
  \bibinfo{pages}{71} (\bibinfo{year}{2005}), \eprint{hep-th/0502005}.

\bibitem[{\citenamefont{Cvetic et~al.}(2004)\citenamefont{Cvetic, Li, and
  Liu}}]{Cvetic:2004ui}
\bibinfo{author}{\bibfnamefont{M.}~\bibnamefont{Cvetic}},
  \bibinfo{author}{\bibfnamefont{T.}~\bibnamefont{Li}}, \bibnamefont{and}
  \bibinfo{author}{\bibfnamefont{T.}~\bibnamefont{Liu}},
  {``}\bibinfo{title}{{Supersymmetric patiSalam models from intersecting
  D6-branes: A Road to the standard model}},{''} \bibinfo{journal}{Nucl.Phys.}
  \textbf{\bibinfo{volume}{B698}}, \bibinfo{pages}{163} (\bibinfo{year}{2004}),
  \eprint{hep-th/0403061}.

\bibitem[{\citenamefont{Cvetic et~al.}(2005)\citenamefont{Cvetic, Langacker,
  Li, and Liu}}]{Cvetic:2004nk}
\bibinfo{author}{\bibfnamefont{M.}~\bibnamefont{Cvetic}},
  \bibinfo{author}{\bibfnamefont{P.}~\bibnamefont{Langacker}},
  \bibinfo{author}{\bibfnamefont{T.-j.} \bibnamefont{Li}}, \bibnamefont{and}
  \bibinfo{author}{\bibfnamefont{T.}~\bibnamefont{Liu}},
  {``}\bibinfo{title}{{D6-brane splitting on type IIA orientifolds}},{''}
  \bibinfo{journal}{Nucl.Phys.} \textbf{\bibinfo{volume}{B709}},
  \bibinfo{pages}{241} (\bibinfo{year}{2005}), \eprint{hep-th/0407178}.

\bibitem[{\citenamefont{Chen et~al.}(2006{\natexlab{a}})\citenamefont{Chen, Li,
  and Nanopoulos}}]{Chen:2005mj}
\bibinfo{author}{\bibfnamefont{C.-M.} \bibnamefont{Chen}},
  \bibinfo{author}{\bibfnamefont{T.}~\bibnamefont{Li}}, \bibnamefont{and}
  \bibinfo{author}{\bibfnamefont{D.~V.} \bibnamefont{Nanopoulos}},
  {``}\bibinfo{title}{{Standard-like model building on Type II
  orientifolds}},{''} \bibinfo{journal}{Nucl.Phys.}
  \textbf{\bibinfo{volume}{B732}}, \bibinfo{pages}{224}
  (\bibinfo{year}{2006}{\natexlab{a}}), \eprint{hep-th/0509059}.

\bibitem[{\citenamefont{Chen et~al.}(2005{\natexlab{a}})\citenamefont{Chen,
  Kraniotis, Mayes, Nanopoulos, and Walker}}]{Chen:2005aba}
\bibinfo{author}{\bibfnamefont{C.~M.} \bibnamefont{Chen}},
  \bibinfo{author}{\bibfnamefont{G.~V.} \bibnamefont{Kraniotis}},
  \bibinfo{author}{\bibfnamefont{V.~E.} \bibnamefont{Mayes}},
  \bibinfo{author}{\bibfnamefont{D.~V.} \bibnamefont{Nanopoulos}},
  \bibnamefont{and} \bibinfo{author}{\bibfnamefont{J.~W.}
  \bibnamefont{Walker}}, {``}\bibinfo{title}{{A Supersymmetric flipped SU(5)
  intersecting brane world}},{''} \bibinfo{journal}{Phys. Lett.}
  \textbf{\bibinfo{volume}{B611}}, \bibinfo{pages}{156}
  (\bibinfo{year}{2005}{\natexlab{a}}), \eprint{hep-th/0501182}.

\bibitem[{\citenamefont{Chen et~al.}(2005{\natexlab{b}})\citenamefont{Chen,
  Kraniotis, Mayes, Nanopoulos, and Walker}}]{Chen:2005mm}
\bibinfo{author}{\bibfnamefont{C.-M.} \bibnamefont{Chen}},
  \bibinfo{author}{\bibfnamefont{G.}~\bibnamefont{Kraniotis}},
  \bibinfo{author}{\bibfnamefont{V.}~\bibnamefont{Mayes}},
  \bibinfo{author}{\bibfnamefont{D.~V.} \bibnamefont{Nanopoulos}},
  \bibnamefont{and} \bibinfo{author}{\bibfnamefont{J.}~\bibnamefont{Walker}},
  {``}\bibinfo{title}{{A K-theory anomaly free supersymmetric flipped SU(5)
  model from intersecting branes}},{''} \bibinfo{journal}{Phys.Lett.}
  \textbf{\bibinfo{volume}{B625}}, \bibinfo{pages}{96}
  (\bibinfo{year}{2005}{\natexlab{b}}), \eprint{hep-th/0507232}.

\bibitem[{\citenamefont{Chen et~al.}(2008{\natexlab{a}})\citenamefont{Chen, Li,
  Mayes, and Nanopoulos}}]{Chen:2007px}
\bibinfo{author}{\bibfnamefont{C.-M.} \bibnamefont{Chen}},
  \bibinfo{author}{\bibfnamefont{T.}~\bibnamefont{Li}},
  \bibinfo{author}{\bibfnamefont{V.}~\bibnamefont{Mayes}}, \bibnamefont{and}
  \bibinfo{author}{\bibfnamefont{D.~V.} \bibnamefont{Nanopoulos}},
  {``}\bibinfo{title}{{A Realistic world from intersecting D6-branes}},{''}
  \bibinfo{journal}{Phys. Lett.} \textbf{\bibinfo{volume}{B665}},
  \bibinfo{pages}{267} (\bibinfo{year}{2008}{\natexlab{a}}),
  \eprint{hep-th/0703280}.

\bibitem[{\citenamefont{Chen et~al.}(2008{\natexlab{b}})\citenamefont{Chen, Li,
  Mayes, and Nanopoulos}}]{Chen:2007zu}
\bibinfo{author}{\bibfnamefont{C.-M.} \bibnamefont{Chen}},
  \bibinfo{author}{\bibfnamefont{T.}~\bibnamefont{Li}},
  \bibinfo{author}{\bibfnamefont{V.}~\bibnamefont{Mayes}}, \bibnamefont{and}
  \bibinfo{author}{\bibfnamefont{D.}~\bibnamefont{Nanopoulos}},
  {``}\bibinfo{title}{{Towards realistic supersymmetric spectra and Yukawa
  textures from intersecting branes}},{''} \bibinfo{journal}{Phys. Rev.}
  \textbf{\bibinfo{volume}{D77}}, \bibinfo{pages}{125023}
  (\bibinfo{year}{2008}{\natexlab{b}}), \eprint{0711.0396}.

\bibitem[{\citenamefont{Maxin et~al.}(2010)\citenamefont{Maxin, Mayes, and
  Nanopoulos}}]{Maxin:2009ez}
\bibinfo{author}{\bibfnamefont{J.~A.} \bibnamefont{Maxin}},
  \bibinfo{author}{\bibfnamefont{V.~E.} \bibnamefont{Mayes}}, \bibnamefont{and}
  \bibinfo{author}{\bibfnamefont{D.~V.} \bibnamefont{Nanopoulos}},
  {``}\bibinfo{title}{{The Search for a Realistic String Model at LHC}},{''}
  \bibinfo{journal}{Phys. Rev. D} \textbf{\bibinfo{volume}{81}},
  \bibinfo{pages}{015008} (\bibinfo{year}{2010}), \eprint{0908.0915}.

\bibitem[{\citenamefont{Li et~al.}(2011{\natexlab{a}})\citenamefont{Li, Maxin,
  Nanopoulos, and Walker}}]{Li:2010rz}
\bibinfo{author}{\bibfnamefont{T.}~\bibnamefont{Li}},
  \bibinfo{author}{\bibfnamefont{J.~A.} \bibnamefont{Maxin}},
  \bibinfo{author}{\bibfnamefont{D.~V.} \bibnamefont{Nanopoulos}},
  \bibnamefont{and} \bibinfo{author}{\bibfnamefont{J.~W.}
  \bibnamefont{Walker}}, {``}\bibinfo{title}{{Dark Matter, Proton Decay and
  Other Phenomenological Constraints in ${\cal F}$-SU(5)}},{''}
  \bibinfo{journal}{Nucl. Phys.} \textbf{\bibinfo{volume}{B848}},
  \bibinfo{pages}{314} (\bibinfo{year}{2011}{\natexlab{a}}),
  \eprint{1003.4186}.

\bibitem[{\citenamefont{Li et~al.}(2011{\natexlab{b}})\citenamefont{Li, Maxin,
  Nanopoulos, and Walker}}]{Li:2010ws}
\bibinfo{author}{\bibfnamefont{T.}~\bibnamefont{Li}},
  \bibinfo{author}{\bibfnamefont{J.~A.} \bibnamefont{Maxin}},
  \bibinfo{author}{\bibfnamefont{D.~V.} \bibnamefont{Nanopoulos}},
  \bibnamefont{and} \bibinfo{author}{\bibfnamefont{J.~W.}
  \bibnamefont{Walker}}, {``}\bibinfo{title}{{The Golden Point of No-Scale and
  No-Parameter ${\cal F}$-$SU(5)$}},{''} \bibinfo{journal}{Phys. Rev.}
  \textbf{\bibinfo{volume}{D83}}, \bibinfo{pages}{056015}
  (\bibinfo{year}{2011}{\natexlab{b}}), \eprint{1007.5100}.

\bibitem[{\citenamefont{Li et~al.}(2011{\natexlab{c}})\citenamefont{Li, Maxin,
  Nanopoulos, and Walker}}]{Li:2010mi}
\bibinfo{author}{\bibfnamefont{T.}~\bibnamefont{Li}},
  \bibinfo{author}{\bibfnamefont{J.~A.} \bibnamefont{Maxin}},
  \bibinfo{author}{\bibfnamefont{D.~V.} \bibnamefont{Nanopoulos}},
  \bibnamefont{and} \bibinfo{author}{\bibfnamefont{J.~W.}
  \bibnamefont{Walker}}, {``}\bibinfo{title}{{The Golden Strip of Correlated
  Top Quark, Gaugino, and Vectorlike Mass In No-Scale, No-Parameter ${\cal
  F}$-$SU(5)$}},{''} \bibinfo{journal}{Phys. Lett.}
  \textbf{\bibinfo{volume}{B699}}, \bibinfo{pages}{164}
  (\bibinfo{year}{2011}{\natexlab{c}}), \eprint{1009.2981}.

\bibitem[{\citenamefont{Li et~al.}(2011{\natexlab{d}})\citenamefont{Li, Maxin,
  Nanopoulos, and Walker}}]{Maxin:2011hy}
\bibinfo{author}{\bibfnamefont{T.}~\bibnamefont{Li}},
  \bibinfo{author}{\bibfnamefont{J.~A.} \bibnamefont{Maxin}},
  \bibinfo{author}{\bibfnamefont{D.~V.} \bibnamefont{Nanopoulos}},
  \bibnamefont{and} \bibinfo{author}{\bibfnamefont{J.~W.}
  \bibnamefont{Walker}}, {``}\bibinfo{title}{{The Ultrahigh jet multiplicity
  signal of stringy no-scale ${\cal F}$-$SU(5)$ at the $\sqrt{s}= 7$ TeV
  LHC}},{''} \bibinfo{journal}{Phys.Rev.} \textbf{\bibinfo{volume}{D84}},
  \bibinfo{pages}{076003} (\bibinfo{year}{2011}{\natexlab{d}}),
  \eprint{1103.4160}.

\bibitem[{\citenamefont{Li et~al.}(2012{\natexlab{a}})\citenamefont{Li, Maxin,
  Nanopoulos, and Walker}}]{Li:2011xua}
\bibinfo{author}{\bibfnamefont{T.}~\bibnamefont{Li}},
  \bibinfo{author}{\bibfnamefont{J.~A.} \bibnamefont{Maxin}},
  \bibinfo{author}{\bibfnamefont{D.~V.} \bibnamefont{Nanopoulos}},
  \bibnamefont{and} \bibinfo{author}{\bibfnamefont{J.~W.}
  \bibnamefont{Walker}}, {``}\bibinfo{title}{{The Unification of Dynamical
  Determination and Bare Minimal Phenomenological Constraints in No-Scale
  ${\cal F}$-$SU(5)$}},{''} \bibinfo{journal}{Phys.Rev.}
  \textbf{\bibinfo{volume}{D85}}, \bibinfo{pages}{056007}
  (\bibinfo{year}{2012}{\natexlab{a}}), \eprint{1105.3988}.

\bibitem[{\citenamefont{Li et~al.}(2012{\natexlab{b}})\citenamefont{Li, Maxin,
  Nanopoulos, and Walker}}]{Li:2011in}
\bibinfo{author}{\bibfnamefont{T.}~\bibnamefont{Li}},
  \bibinfo{author}{\bibfnamefont{J.~A.} \bibnamefont{Maxin}},
  \bibinfo{author}{\bibfnamefont{D.~V.} \bibnamefont{Nanopoulos}},
  \bibnamefont{and} \bibinfo{author}{\bibfnamefont{J.~W.}
  \bibnamefont{Walker}}, {``}\bibinfo{title}{{The Race for Supersymmetric Dark
  Matter at XENON100 and the LHC: Stringy Correlations from No-Scale ${\cal
  F}$-$SU(5)$}},{''} \bibinfo{journal}{JHEP} \textbf{\bibinfo{volume}{1212}},
  \bibinfo{pages}{017} (\bibinfo{year}{2012}{\natexlab{b}}),
  \eprint{1106.1165}.

\bibitem[{\citenamefont{Li et~al.}(2012{\natexlab{c}})\citenamefont{Li, Maxin,
  Nanopoulos, and Walker}}]{Li:2011gh}
\bibinfo{author}{\bibfnamefont{T.}~\bibnamefont{Li}},
  \bibinfo{author}{\bibfnamefont{J.~A.} \bibnamefont{Maxin}},
  \bibinfo{author}{\bibfnamefont{D.~V.} \bibnamefont{Nanopoulos}},
  \bibnamefont{and} \bibinfo{author}{\bibfnamefont{J.~W.}
  \bibnamefont{Walker}}, {``}\bibinfo{title}{{A Two-Tiered Correlation of Dark
  Matter with Missing Transverse Energy: Reconstructing the Lightest
  Supersymmetric Particle Mass at the LHC}},{''} \bibinfo{journal}{JHEP}
  \textbf{\bibinfo{volume}{02}}, \bibinfo{pages}{129}
  (\bibinfo{year}{2012}{\natexlab{c}}), \eprint{1107.2375}.

\bibitem[{\citenamefont{Li et~al.}(2012{\natexlab{d}})\citenamefont{Li, Maxin,
  Nanopoulos, and Walker}}]{Li:2011ab}
\bibinfo{author}{\bibfnamefont{T.}~\bibnamefont{Li}},
  \bibinfo{author}{\bibfnamefont{J.~A.} \bibnamefont{Maxin}},
  \bibinfo{author}{\bibfnamefont{D.~V.} \bibnamefont{Nanopoulos}},
  \bibnamefont{and} \bibinfo{author}{\bibfnamefont{J.~W.}
  \bibnamefont{Walker}}, {``}\bibinfo{title}{{A Higgs Mass Shift to 125 GeV and
  A Multi-Jet Supersymmetry Signal: Miracle of the Flippons at the $\sqrt{s} =
  7$~TeV LHC}},{''} \bibinfo{journal}{Phys.Lett.}
  \textbf{\bibinfo{volume}{B710}}, \bibinfo{pages}{207}
  (\bibinfo{year}{2012}{\natexlab{d}}), \eprint{1112.3024}.

\bibitem[{\citenamefont{Li et~al.}(2013)\citenamefont{Li, Maxin, Nanopoulos,
  and Walker}}]{Li:2013naa}
\bibinfo{author}{\bibfnamefont{T.}~\bibnamefont{Li}},
  \bibinfo{author}{\bibfnamefont{J.~A.} \bibnamefont{Maxin}},
  \bibinfo{author}{\bibfnamefont{D.~V.} \bibnamefont{Nanopoulos}},
  \bibnamefont{and} \bibinfo{author}{\bibfnamefont{J.~W.}
  \bibnamefont{Walker}}, {``}\bibinfo{title}{{No-Scale ${\cal F}$-$SU(5)$ in
  the Light of LHC, Planck and XENON}},{''} \bibinfo{journal}{Jour.Phys.}
  \textbf{\bibinfo{volume}{G40}}, \bibinfo{pages}{115002}
  (\bibinfo{year}{2013}), \eprint{1305.1846}.

\bibitem[{\citenamefont{Li et~al.}(2014)\citenamefont{Li, Maxin, Nanopoulos,
  and Walker}}]{Li:2013mwa}
\bibinfo{author}{\bibfnamefont{T.}~\bibnamefont{Li}},
  \bibinfo{author}{\bibfnamefont{J.~A.} \bibnamefont{Maxin}},
  \bibinfo{author}{\bibfnamefont{D.~V.} \bibnamefont{Nanopoulos}},
  \bibnamefont{and} \bibinfo{author}{\bibfnamefont{J.~W.}
  \bibnamefont{Walker}}, {``}\bibinfo{title}{{Testing No-Scale Supergravity
  with the Fermi Space Telescope LAT}},{''} \bibinfo{journal}{J. Phys.}
  \textbf{\bibinfo{volume}{G41}}, \bibinfo{pages}{055006}
  (\bibinfo{year}{2014}), \eprint{1311.1164}.

\bibitem[{\citenamefont{Li et~al.}(2017)\citenamefont{Li, Maxin, and
  Nanopoulos}}]{Li:2016bww}
\bibinfo{author}{\bibfnamefont{T.}~\bibnamefont{Li}},
  \bibinfo{author}{\bibfnamefont{J.~A.} \bibnamefont{Maxin}}, \bibnamefont{and}
  \bibinfo{author}{\bibfnamefont{D.~V.} \bibnamefont{Nanopoulos}},
  {``}\bibinfo{title}{{The return of the King: No-Scale ${\cal
  F}$-$SU(5)$}},{''} \bibinfo{journal}{Phys. Lett.}
  \textbf{\bibinfo{volume}{B764}}, \bibinfo{pages}{167} (\bibinfo{year}{2017}),
  \eprint{1609.06294}.

\bibitem[{\citenamefont{Ford et~al.}(2019)\citenamefont{Ford, Li, Maxin, and
  Nanopoulos}}]{Ford:2019kzv}
\bibinfo{author}{\bibfnamefont{T.}~\bibnamefont{Ford}},
  \bibinfo{author}{\bibfnamefont{T.}~\bibnamefont{Li}},
  \bibinfo{author}{\bibfnamefont{J.~A.} \bibnamefont{Maxin}}, \bibnamefont{and}
  \bibinfo{author}{\bibfnamefont{D.~V.} \bibnamefont{Nanopoulos}},
  {``}\bibinfo{title}{{The heavy gluino in natural no-scale
  $\cal{F}$-$SU$(5)}},{''} \bibinfo{journal}{Phys. Lett. B}
  \textbf{\bibinfo{volume}{799}}, \bibinfo{pages}{135038}
  (\bibinfo{year}{2019}), \eprint{1908.06149}.

\bibitem[{\citenamefont{Ellis et~al.}(2019{\natexlab{a}})\citenamefont{Ellis,
  Garcia, Nagata, Nanopoulos, and Olive}}]{Ellis:2019jha}
\bibinfo{author}{\bibfnamefont{J.}~\bibnamefont{Ellis}},
  \bibinfo{author}{\bibfnamefont{M.~A.~G.} \bibnamefont{Garcia}},
  \bibinfo{author}{\bibfnamefont{N.}~\bibnamefont{Nagata}},
  \bibinfo{author}{\bibfnamefont{D.~V.} \bibnamefont{Nanopoulos}},
  \bibnamefont{and} \bibinfo{author}{\bibfnamefont{K.~A.} \bibnamefont{Olive}},
  {``}\bibinfo{title}{{Cosmology with a master coupling in flipped SU(5)
  $\times$ U(1): the $\lambda_6$ universe}},{''} \bibinfo{journal}{Phys. Lett.
  B} \textbf{\bibinfo{volume}{797}}, \bibinfo{pages}{134864}
  (\bibinfo{year}{2019}{\natexlab{a}}), \eprint{1906.08483}.

\bibitem[{\citenamefont{Ellis et~al.}(2019{\natexlab{b}})\citenamefont{Ellis,
  Nanopoulos, Olive, and Verner}}]{Ellis:2019bmm}
\bibinfo{author}{\bibfnamefont{J.}~\bibnamefont{Ellis}},
  \bibinfo{author}{\bibfnamefont{D.~V.} \bibnamefont{Nanopoulos}},
  \bibinfo{author}{\bibfnamefont{K.~A.} \bibnamefont{Olive}}, \bibnamefont{and}
  \bibinfo{author}{\bibfnamefont{S.}~\bibnamefont{Verner}},
  {``}\bibinfo{title}{{Unified No-Scale Attractors}},{''}
  \bibinfo{journal}{JCAP} \textbf{\bibinfo{volume}{09}}, \bibinfo{pages}{040}
  (\bibinfo{year}{2019}{\natexlab{b}}), \eprint{1906.10176}.

\bibitem[{\citenamefont{Ellis et~al.}(2019{\natexlab{c}})\citenamefont{Ellis,
  Nagaraj, Nanopoulos, Olive, and Verner}}]{Ellis:2019hps}
\bibinfo{author}{\bibfnamefont{J.}~\bibnamefont{Ellis}},
  \bibinfo{author}{\bibfnamefont{B.}~\bibnamefont{Nagaraj}},
  \bibinfo{author}{\bibfnamefont{D.~V.} \bibnamefont{Nanopoulos}},
  \bibinfo{author}{\bibfnamefont{K.~A.} \bibnamefont{Olive}}, \bibnamefont{and}
  \bibinfo{author}{\bibfnamefont{S.}~\bibnamefont{Verner}},
  {``}\bibinfo{title}{{From Minkowski to de Sitter in Multifield No-Scale
  Models}},{''} \bibinfo{journal}{JHEP} \textbf{\bibinfo{volume}{10}},
  \bibinfo{pages}{161} (\bibinfo{year}{2019}{\natexlab{c}}),
  \eprint{1907.09123}.

\bibitem[{\citenamefont{Ellis et~al.}(2020{\natexlab{a}})\citenamefont{Ellis,
  Garcia, Nagata, Nanopoulos, and Olive}}]{Ellis:2019opr}
\bibinfo{author}{\bibfnamefont{J.}~\bibnamefont{Ellis}},
  \bibinfo{author}{\bibfnamefont{M.~A.~G.} \bibnamefont{Garcia}},
  \bibinfo{author}{\bibfnamefont{N.}~\bibnamefont{Nagata}},
  \bibinfo{author}{\bibfnamefont{D.~V.} \bibnamefont{Nanopoulos}},
  \bibnamefont{and} \bibinfo{author}{\bibfnamefont{K.~A.} \bibnamefont{Olive}},
  {``}\bibinfo{title}{{Superstring-Inspired Particle Cosmology: Inflation,
  Neutrino Masses, Leptogenesis, Dark Matter \& the SUSY Scale}},{''}
  \bibinfo{journal}{JCAP} \textbf{\bibinfo{volume}{01}}, \bibinfo{pages}{035}
  (\bibinfo{year}{2020}{\natexlab{a}}), \eprint{1910.11755}.

\bibitem[{\citenamefont{Ellis et~al.}(2020{\natexlab{b}})\citenamefont{Ellis,
  Garcia, Nagata, Nanopoulos, and Olive}}]{Ellis:2020qad}
\bibinfo{author}{\bibfnamefont{J.}~\bibnamefont{Ellis}},
  \bibinfo{author}{\bibfnamefont{M.~A.~G.} \bibnamefont{Garcia}},
  \bibinfo{author}{\bibfnamefont{N.}~\bibnamefont{Nagata}},
  \bibinfo{author}{\bibfnamefont{D.~V.} \bibnamefont{Nanopoulos}},
  \bibnamefont{and} \bibinfo{author}{\bibfnamefont{K.~A.} \bibnamefont{Olive}},
  {``}\bibinfo{title}{{Proton Decay: Flipped vs Unflipped SU(5)}},{''}
  \bibinfo{journal}{JHEP} \textbf{\bibinfo{volume}{05}}, \bibinfo{pages}{021}
  (\bibinfo{year}{2020}{\natexlab{b}}), \eprint{2003.03285}.

\bibitem[{\citenamefont{Ellis et~al.}(2020{\natexlab{c}})\citenamefont{Ellis,
  Nanopoulos, Olive, and Verner}}]{Ellis:2020xmk}
\bibinfo{author}{\bibfnamefont{J.}~\bibnamefont{Ellis}},
  \bibinfo{author}{\bibfnamefont{D.~V.} \bibnamefont{Nanopoulos}},
  \bibinfo{author}{\bibfnamefont{K.~A.} \bibnamefont{Olive}}, \bibnamefont{and}
  \bibinfo{author}{\bibfnamefont{S.}~\bibnamefont{Verner}},
  {``}\bibinfo{title}{{Phenomenology and Cosmology of No-Scale Attractor Models
  of Inflation}},{''} \bibinfo{journal}{JCAP} \textbf{\bibinfo{volume}{08}},
  \bibinfo{pages}{037} (\bibinfo{year}{2020}{\natexlab{c}}),
  \eprint{2004.00643}.

\bibitem[{\citenamefont{Ellis et~al.}(2020{\natexlab{d}})\citenamefont{Ellis,
  Mavromatos, and Nanopoulos}}]{Ellis:2020nnp}
\bibinfo{author}{\bibfnamefont{J.}~\bibnamefont{Ellis}},
  \bibinfo{author}{\bibfnamefont{N.~E.} \bibnamefont{Mavromatos}},
  \bibnamefont{and} \bibinfo{author}{\bibfnamefont{D.~V.}
  \bibnamefont{Nanopoulos}}, {``}\bibinfo{title}{{Supercritical String
  Cosmology drains the Swampland}},{''} \bibinfo{journal}{Phys. Rev. D}
  \textbf{\bibinfo{volume}{102}}, \bibinfo{pages}{046015}
  (\bibinfo{year}{2020}{\natexlab{d}}), \eprint{2006.06430}.

\bibitem[{\citenamefont{Nanopoulos et~al.}(2020)\citenamefont{Nanopoulos,
  Spanos, and Stamou}}]{Nanopoulos:2020nnh}
\bibinfo{author}{\bibfnamefont{D.~V.} \bibnamefont{Nanopoulos}},
  \bibinfo{author}{\bibfnamefont{V.~C.} \bibnamefont{Spanos}},
  \bibnamefont{and} \bibinfo{author}{\bibfnamefont{I.~D.}
  \bibnamefont{Stamou}}, {``}\bibinfo{title}{{Primordial Black Holes from
  No-Scale Supergravity}},{''} \bibinfo{journal}{Phys. Rev. D}
  \textbf{\bibinfo{volume}{102}}, \bibinfo{pages}{083536}
  (\bibinfo{year}{2020}), \eprint{2008.01457}.

\bibitem[{\citenamefont{Ellis et~al.}(2021)\citenamefont{Ellis, Nanopoulos,
  Olive, and Verner}}]{Ellis:2020krl}
\bibinfo{author}{\bibfnamefont{J.}~\bibnamefont{Ellis}},
  \bibinfo{author}{\bibfnamefont{D.~V.} \bibnamefont{Nanopoulos}},
  \bibinfo{author}{\bibfnamefont{K.~A.} \bibnamefont{Olive}}, \bibnamefont{and}
  \bibinfo{author}{\bibfnamefont{S.}~\bibnamefont{Verner}},
  {``}\bibinfo{title}{{Non-Oscillatory No-Scale Inflation}},{''}
  \bibinfo{journal}{JCAP} \textbf{\bibinfo{volume}{03}}, \bibinfo{pages}{052}
  (\bibinfo{year}{2021}), \eprint{2008.09099}.

\bibitem[{\citenamefont{Antoniadis et~al.}(2021)\citenamefont{Antoniadis,
  Nanopoulos, and Rizos}}]{Antoniadis:2020txn}
\bibinfo{author}{\bibfnamefont{I.}~\bibnamefont{Antoniadis}},
  \bibinfo{author}{\bibfnamefont{D.~V.} \bibnamefont{Nanopoulos}},
  \bibnamefont{and} \bibinfo{author}{\bibfnamefont{J.}~\bibnamefont{Rizos}},
  {``}\bibinfo{title}{{Cosmology of the string derived flipped $SU(5)$}},{''}
  \bibinfo{journal}{JCAP} \textbf{\bibinfo{volume}{03}}, \bibinfo{pages}{017}
  (\bibinfo{year}{2021}), \eprint{2011.09396}.

\bibitem[{\citenamefont{Leggett et~al.}(2015)\citenamefont{Leggett, Li, Maxin,
  Nanopoulos, and Walker}}]{Leggett:2014hha}
\bibinfo{author}{\bibfnamefont{T.}~\bibnamefont{Leggett}},
  \bibinfo{author}{\bibfnamefont{T.}~\bibnamefont{Li}},
  \bibinfo{author}{\bibfnamefont{J.~A.} \bibnamefont{Maxin}},
  \bibinfo{author}{\bibfnamefont{D.~V.} \bibnamefont{Nanopoulos}},
  \bibnamefont{and} \bibinfo{author}{\bibfnamefont{J.~W.}
  \bibnamefont{Walker}}, {``}\bibinfo{title}{{Confronting Electroweak
  Fine-tuning with No-Scale Supergravity}},{''} \bibinfo{journal}{Phys.Lett.}
  \textbf{\bibinfo{volume}{B740}}, \bibinfo{pages}{66} (\bibinfo{year}{2015}),
  \eprint{1408.4459}.

\bibitem[{\citenamefont{De~Benedetti et~al.}(2018)\citenamefont{De~Benedetti,
  Li, Li, Lux, Maxin, and Nanopoulos}}]{DeBenedetti:2018fxa}
\bibinfo{author}{\bibfnamefont{R.}~\bibnamefont{De~Benedetti}},
  \bibinfo{author}{\bibfnamefont{C.}~\bibnamefont{Li}},
  \bibinfo{author}{\bibfnamefont{T.}~\bibnamefont{Li}},
  \bibinfo{author}{\bibfnamefont{A.}~\bibnamefont{Lux}},
  \bibinfo{author}{\bibfnamefont{J.~A.} \bibnamefont{Maxin}}, \bibnamefont{and}
  \bibinfo{author}{\bibfnamefont{D.~V.} \bibnamefont{Nanopoulos}},
  {``}\bibinfo{title}{{Inspiration from intersecting D-branes: general
  supersymmetry breaking soft terms in no-scale $\mathcal{F}$ -SU(5)}},{''}
  \bibinfo{journal}{Eur. Phys. J.} \textbf{\bibinfo{volume}{C78}},
  \bibinfo{pages}{958} (\bibinfo{year}{2018}), \eprint{1809.09695}.

\bibitem[{\citenamefont{De~Benedetti et~al.}(2019)\citenamefont{De~Benedetti,
  Li, Maxin, and Nanopoulos}}]{DeBenedetti:2019hrk}
\bibinfo{author}{\bibfnamefont{R.}~\bibnamefont{De~Benedetti}},
  \bibinfo{author}{\bibfnamefont{T.}~\bibnamefont{Li}},
  \bibinfo{author}{\bibfnamefont{J.~A.} \bibnamefont{Maxin}}, \bibnamefont{and}
  \bibinfo{author}{\bibfnamefont{D.~V.} \bibnamefont{Nanopoulos}},
  {``}\bibinfo{title}{{Naturalness in D-brane inspired models}},{''}
  \bibinfo{journal}{JHEP} \textbf{\bibinfo{volume}{07}}, \bibinfo{pages}{048}
  (\bibinfo{year}{2019}), \eprint{1904.10809}.

\bibitem[{\citenamefont{Barr}(1982)}]{Barr:1981qv}
\bibinfo{author}{\bibfnamefont{S.~M.} \bibnamefont{Barr}},
  {``}\bibinfo{title}{{A New Symmetry Breaking Pattern for $SO(10)$ and Proton
  Decay}},{''} \bibinfo{journal}{Phys. Lett.} \textbf{\bibinfo{volume}{B112}},
  \bibinfo{pages}{219} (\bibinfo{year}{1982}).

\bibitem[{\citenamefont{Derendinger et~al.}(1984)\citenamefont{Derendinger,
  Kim, and Nanopoulos}}]{Derendinger:1983aj}
\bibinfo{author}{\bibfnamefont{J.~P.} \bibnamefont{Derendinger}},
  \bibinfo{author}{\bibfnamefont{J.~E.} \bibnamefont{Kim}}, \bibnamefont{and}
  \bibinfo{author}{\bibfnamefont{D.~V.} \bibnamefont{Nanopoulos}},
  {``}\bibinfo{title}{{Anti-$SU(5)$}},{''} \bibinfo{journal}{Phys. Lett.}
  \textbf{\bibinfo{volume}{B139}}, \bibinfo{pages}{170} (\bibinfo{year}{1984}).

\bibitem[{\citenamefont{Antoniadis et~al.}(1987)\citenamefont{Antoniadis,
  Ellis, Hagelin, and Nanopoulos}}]{Antoniadis:1987dx}
\bibinfo{author}{\bibfnamefont{I.}~\bibnamefont{Antoniadis}},
  \bibinfo{author}{\bibfnamefont{J.~R.} \bibnamefont{Ellis}},
  \bibinfo{author}{\bibfnamefont{J.~S.} \bibnamefont{Hagelin}},
  \bibnamefont{and} \bibinfo{author}{\bibfnamefont{D.~V.}
  \bibnamefont{Nanopoulos}}, {``}\bibinfo{title}{{Supersymmetric Flipped
  $SU(5)$ Revitalized}},{''} \bibinfo{journal}{Phys. Lett.}
  \textbf{\bibinfo{volume}{B194}}, \bibinfo{pages}{231} (\bibinfo{year}{1987}).

\bibitem[{\citenamefont{Harnik et~al.}(2005)\citenamefont{Harnik, Larson,
  Murayama, and Thormeier}}]{Harnik:2004yp}
\bibinfo{author}{\bibfnamefont{R.}~\bibnamefont{Harnik}},
  \bibinfo{author}{\bibfnamefont{D.~T.} \bibnamefont{Larson}},
  \bibinfo{author}{\bibfnamefont{H.}~\bibnamefont{Murayama}}, \bibnamefont{and}
  \bibinfo{author}{\bibfnamefont{M.}~\bibnamefont{Thormeier}},
  {``}\bibinfo{title}{{Probing the Planck scale with proton decay}},{''}
  \bibinfo{journal}{Nucl.Phys.} \textbf{\bibinfo{volume}{B706}},
  \bibinfo{pages}{372} (\bibinfo{year}{2005}), \eprint{hep-ph/0404260}.

\bibitem[{\citenamefont{Jiang et~al.}(2007)\citenamefont{Jiang, Li, and
  Nanopoulos}}]{Jiang:2006hf}
\bibinfo{author}{\bibfnamefont{J.}~\bibnamefont{Jiang}},
  \bibinfo{author}{\bibfnamefont{T.}~\bibnamefont{Li}}, \bibnamefont{and}
  \bibinfo{author}{\bibfnamefont{D.~V.} \bibnamefont{Nanopoulos}},
  {``}\bibinfo{title}{{Testable Flipped $SU(5) \times U(1)_X$ Models}},{''}
  \bibinfo{journal}{Nucl. Phys.} \textbf{\bibinfo{volume}{B772}},
  \bibinfo{pages}{49} (\bibinfo{year}{2007}), \eprint{hep-ph/0610054}.

\bibitem[{\citenamefont{Jiang et~al.}(2009)\citenamefont{Jiang, Li, Nanopoulos,
  and Xie}}]{Jiang:2008yf}
\bibinfo{author}{\bibfnamefont{J.}~\bibnamefont{Jiang}},
  \bibinfo{author}{\bibfnamefont{T.}~\bibnamefont{Li}},
  \bibinfo{author}{\bibfnamefont{D.~V.} \bibnamefont{Nanopoulos}},
  \bibnamefont{and} \bibinfo{author}{\bibfnamefont{D.}~\bibnamefont{Xie}},
  {``}\bibinfo{title}{{F-SU(5)}},{''} \bibinfo{journal}{Phys. Lett.}
  \textbf{\bibinfo{volume}{B677}}, \bibinfo{pages}{322} (\bibinfo{year}{2009}),
  \eprint{0811.2807}.

\bibitem[{\citenamefont{Jiang et~al.}(2010)\citenamefont{Jiang, Li, Nanopoulos,
  and Xie}}]{Jiang:2009za}
\bibinfo{author}{\bibfnamefont{J.}~\bibnamefont{Jiang}},
  \bibinfo{author}{\bibfnamefont{T.}~\bibnamefont{Li}},
  \bibinfo{author}{\bibfnamefont{D.~V.} \bibnamefont{Nanopoulos}},
  \bibnamefont{and} \bibinfo{author}{\bibfnamefont{D.}~\bibnamefont{Xie}},
  {``}\bibinfo{title}{{Flipped $SU(5) \times U(1)_X$ Models from
  F-Theory}},{''} \bibinfo{journal}{Nucl. Phys.}
  \textbf{\bibinfo{volume}{B830}}, \bibinfo{pages}{195} (\bibinfo{year}{2010}),
  \eprint{0905.3394}.

\bibitem[{\citenamefont{Chen et~al.}(2006{\natexlab{b}})\citenamefont{Chen, Li,
  and Nanopoulos}}]{Chen:2006ip}
\bibinfo{author}{\bibfnamefont{C.-M.} \bibnamefont{Chen}},
  \bibinfo{author}{\bibfnamefont{T.}~\bibnamefont{Li}}, \bibnamefont{and}
  \bibinfo{author}{\bibfnamefont{D.~V.} \bibnamefont{Nanopoulos}},
  {``}\bibinfo{title}{{Flipped and unflipped SU(5) as type IIA flux
  vacua}},{''} \bibinfo{journal}{Nucl. Phys.} \textbf{\bibinfo{volume}{B751}},
  \bibinfo{pages}{260} (\bibinfo{year}{2006}{\natexlab{b}}),
  \eprint{hep-th/0604107}.

\bibitem[{\citenamefont{Li and Nanopoulos}(2010)}]{Li:2010xr}
\bibinfo{author}{\bibfnamefont{T.}~\bibnamefont{Li}} \bibnamefont{and}
  \bibinfo{author}{\bibfnamefont{D.~V.} \bibnamefont{Nanopoulos}},
  {``}\bibinfo{title}{{Generalizing Minimal Supergravity}},{''}
  \bibinfo{journal}{Phys.Lett.} \textbf{\bibinfo{volume}{B692}},
  \bibinfo{pages}{121} (\bibinfo{year}{2010}), \eprint{1002.4183}.

\bibitem[{\citenamefont{Balazs et~al.}(2010)\citenamefont{Balazs, Li,
  Nanopoulos, and Wang}}]{Balazs:2010ha}
\bibinfo{author}{\bibfnamefont{C.}~\bibnamefont{Balazs}},
  \bibinfo{author}{\bibfnamefont{T.}~\bibnamefont{Li}},
  \bibinfo{author}{\bibfnamefont{D.~V.} \bibnamefont{Nanopoulos}},
  \bibnamefont{and} \bibinfo{author}{\bibfnamefont{F.}~\bibnamefont{Wang}},
  {``}\bibinfo{title}{{Supersymmetry Breaking Scalar Masses and Trilinear Soft
  Terms in Generalized Minimal Supergravity}},{''} \bibinfo{journal}{JHEP}
  \textbf{\bibinfo{volume}{09}}, \bibinfo{pages}{003} (\bibinfo{year}{2010}),
  \eprint{1006.5559}.

\bibitem[{\citenamefont{Belanger et~al.}(2009)\citenamefont{Belanger, Boudjema,
  Pukhov, and Semenov}}]{Belanger:2008sj}
\bibinfo{author}{\bibfnamefont{G.}~\bibnamefont{Belanger}},
  \bibinfo{author}{\bibfnamefont{F.}~\bibnamefont{Boudjema}},
  \bibinfo{author}{\bibfnamefont{A.}~\bibnamefont{Pukhov}}, \bibnamefont{and}
  \bibinfo{author}{\bibfnamefont{A.}~\bibnamefont{Semenov}},
  {``}\bibinfo{title}{{Dark matter direct detection rate in a generic model
  with micrOMEGAs2.1}},{''} \bibinfo{journal}{Comput. Phys. Commun.}
  \textbf{\bibinfo{volume}{180}}, \bibinfo{pages}{747} (\bibinfo{year}{2009}),
  \eprint{0803.2360}.

\bibitem[{\citenamefont{Djouadi et~al.}(2007)\citenamefont{Djouadi, Kneur, and
  Moultaka}}]{Djouadi:2002ze}
\bibinfo{author}{\bibfnamefont{A.}~\bibnamefont{Djouadi}},
  \bibinfo{author}{\bibfnamefont{J.-L.} \bibnamefont{Kneur}}, \bibnamefont{and}
  \bibinfo{author}{\bibfnamefont{G.}~\bibnamefont{Moultaka}},
  {``}\bibinfo{title}{{SuSpect: A Fortran code for the supersymmetric and Higgs
  particle spectrum in the MSSM}},{''} \bibinfo{journal}{Comput. Phys. Commun.}
  \textbf{\bibinfo{volume}{176}}, \bibinfo{pages}{426} (\bibinfo{year}{2007}),
  \eprint{hep-ph/0211331}.

\bibitem[{\citenamefont{Patrignani et~al.}(2016)}]{Olive:2016xmw}
\bibinfo{author}{\bibfnamefont{C.}~\bibnamefont{Patrignani}}
  \bibnamefont{et~al.} (\bibinfo{collaboration}{Particle Data Group}),
  {``}\bibinfo{title}{{Review of Particle Physics}},{''}
  \bibinfo{journal}{Chin. Phys.} \textbf{\bibinfo{volume}{C40}},
  \bibinfo{pages}{100001} (\bibinfo{year}{2016}).

\bibitem[{\citenamefont{Athron et~al.}(2016)\citenamefont{Athron, Bach,
  Fargnoli, Gnendiger, Greifenhagen, Park, Pa\ss{}ehr, St\"ockinger,
  St\"ockinger-Kim, and Voigt}}]{Athron:2015rva}
\bibinfo{author}{\bibfnamefont{P.}~\bibnamefont{Athron}},
  \bibinfo{author}{\bibfnamefont{M.}~\bibnamefont{Bach}},
  \bibinfo{author}{\bibfnamefont{H.~G.} \bibnamefont{Fargnoli}},
  \bibinfo{author}{\bibfnamefont{C.}~\bibnamefont{Gnendiger}},
  \bibinfo{author}{\bibfnamefont{R.}~\bibnamefont{Greifenhagen}},
  \bibinfo{author}{\bibfnamefont{J.-h.} \bibnamefont{Park}},
  \bibinfo{author}{\bibfnamefont{S.}~\bibnamefont{Pa\ss{}ehr}},
  \bibinfo{author}{\bibfnamefont{D.}~\bibnamefont{St\"ockinger}},
  \bibinfo{author}{\bibfnamefont{H.}~\bibnamefont{St\"ockinger-Kim}},
  \bibnamefont{and} \bibinfo{author}{\bibfnamefont{A.}~\bibnamefont{Voigt}},
  {``}\bibinfo{title}{{GM2Calc: Precise MSSM prediction for $(g - 2)$ of the
  muon}},{''} \bibinfo{journal}{Eur. Phys. J. C} \textbf{\bibinfo{volume}{76}},
  \bibinfo{pages}{62} (\bibinfo{year}{2016}), \eprint{1510.08071}.

\bibitem[{\citenamefont{Aad et~al.}(2020{\natexlab{a}})}]{Aad:2019qnd}
\bibinfo{author}{\bibfnamefont{G.}~\bibnamefont{Aad}} \bibnamefont{et~al.}
  (\bibinfo{collaboration}{ATLAS}), {``}\bibinfo{title}{{Searches for
  electroweak production of supersymmetric particles with compressed mass
  spectra in $\sqrt{s}=$ 13 TeV $pp$ collisions with the ATLAS detector}},{''}
  \bibinfo{journal}{Phys. Rev. D} \textbf{\bibinfo{volume}{101}},
  \bibinfo{pages}{052005} (\bibinfo{year}{2020}{\natexlab{a}}),
  \eprint{1911.12606}.

\bibitem[{CMS(2021{\natexlab{a}})}]{CMS:ew}
{``}\bibinfo{title}{{Search for physics beyond the standard model in final
  states with two or three soft leptons and missing transverse momentum in
  proton-proton collisions at 13 TeV}},{''}
  (\bibinfo{year}{2021}{\natexlab{a}}), \eprint{CMS-PAS-SUS-18-004}.

\bibitem[{\citenamefont{Amhis
  et~al.}(2012)}]{HeavyFlavorAveragingGroup:2012zzm}
\bibinfo{author}{\bibfnamefont{Y.}~\bibnamefont{Amhis}} \bibnamefont{et~al.}
  (\bibinfo{collaboration}{Heavy Flavor Averaging Group}),
  {``}\bibinfo{title}{{Averages of B-Hadron, C-Hadron, and tau-lepton
  properties as of early 2012}},{''} (\bibinfo{year}{2012}),
  \eprint{1207.1158}.

\bibitem[{\citenamefont{Aaij et~al.}(2012)}]{:2012ct}
\bibinfo{author}{\bibfnamefont{R.}~\bibnamefont{Aaij}} \bibnamefont{et~al.}
  (\bibinfo{collaboration}{LHCb Collaboration}), {``}\bibinfo{title}{{First
  evidence for the decay $B_s^0 \to \mu^+ \mu^-$}},{''} (\bibinfo{year}{2012}),
  \eprint{1211.2674}.

\bibitem[{\citenamefont{Hinshaw et~al.}(2013)}]{Hinshaw:2012aka}
\bibinfo{author}{\bibfnamefont{G.}~\bibnamefont{Hinshaw}} \bibnamefont{et~al.}
  (\bibinfo{collaboration}{WMAP}), {``}\bibinfo{title}{{Nine-Year Wilkinson
  Microwave Anisotropy Probe (WMAP) Observations: Cosmological Parameter
  Results}},{''} \bibinfo{journal}{Astrophys. J. Suppl.}
  \textbf{\bibinfo{volume}{208}}, \bibinfo{pages}{19} (\bibinfo{year}{2013}),
  \eprint{1212.5226}.

\bibitem[{\citenamefont{Ade et~al.}(2016)}]{Ade:2015xua}
\bibinfo{author}{\bibfnamefont{P.~A.~R.} \bibnamefont{Ade}}
  \bibnamefont{et~al.} (\bibinfo{collaboration}{Planck}),
  {``}\bibinfo{title}{{Planck 2015 results. XIII. Cosmological
  parameters}},{''} \bibinfo{journal}{Astron. Astrophys.}
  \textbf{\bibinfo{volume}{594}}, \bibinfo{pages}{A13} (\bibinfo{year}{2016}),
  \eprint{1502.01589}.

\bibitem[{\citenamefont{Aghanim et~al.}(2018)}]{Aghanim:2018eyx}
\bibinfo{author}{\bibfnamefont{N.}~\bibnamefont{Aghanim}} \bibnamefont{et~al.}
  (\bibinfo{collaboration}{Planck}), {``}\bibinfo{title}{{Planck 2018 results.
  VI. Cosmological parameters}},{''} (\bibinfo{year}{2018}),
  \eprint{1807.06209}.

\bibitem[{\citenamefont{Takhistov}(2016)}]{Takhistov:2016eqm}
\bibinfo{author}{\bibfnamefont{V.}~\bibnamefont{Takhistov}}
  (\bibinfo{collaboration}{Super-Kamiokande}), in
  \emph{\bibinfo{booktitle}{{Proceedings, 51st Rencontres de Moriond on
  Electroweak Interactions and Unified Theories: La Thuile, Italy, March 12-19,
  2016}}} (\bibinfo{year}{2016}), pp. \bibinfo{pages}{437--444},
  \eprint{1605.03235}.

\bibitem[{\citenamefont{Aad et~al.}(2012)}]{Aad:2012tfa}
\bibinfo{author}{\bibfnamefont{G.}~\bibnamefont{Aad}} \bibnamefont{et~al.}
  (\bibinfo{collaboration}{ATLAS}), {``}\bibinfo{title}{{Observation of a new
  particle in the search for the Standard Model Higgs boson with the ATLAS
  detector at the LHC}},{''} \bibinfo{journal}{Phys. Lett.}
  \textbf{\bibinfo{volume}{B716}}, \bibinfo{pages}{1} (\bibinfo{year}{2012}),
  \eprint{1207.7214}.

\bibitem[{\citenamefont{Chatrchyan et~al.}(2012)}]{Chatrchyan:2012xdj}
\bibinfo{author}{\bibfnamefont{S.}~\bibnamefont{Chatrchyan}}
  \bibnamefont{et~al.} (\bibinfo{collaboration}{CMS}),
  {``}\bibinfo{title}{{Observation of a new boson at a mass of 125 GeV with the
  CMS experiment at the LHC}},{''} \bibinfo{journal}{Phys. Lett.}
  \textbf{\bibinfo{volume}{B716}}, \bibinfo{pages}{30} (\bibinfo{year}{2012}),
  \eprint{1207.7235}.

\bibitem[{\citenamefont{Huo et~al.}(2012)\citenamefont{Huo, Li, Nanopoulos, and
  Tong}}]{Huo:2011zt}
\bibinfo{author}{\bibfnamefont{Y.}~\bibnamefont{Huo}},
  \bibinfo{author}{\bibfnamefont{T.}~\bibnamefont{Li}},
  \bibinfo{author}{\bibfnamefont{D.~V.} \bibnamefont{Nanopoulos}},
  \bibnamefont{and} \bibinfo{author}{\bibfnamefont{C.}~\bibnamefont{Tong}},
  {``}\bibinfo{title}{{The Lightest CP-Even Higgs Boson Mass in the Testable
  Flipped $SU(5) \times U(1)_X$ Models from F-Theory}},{''}
  \bibinfo{journal}{Phys.Rev.} \textbf{\bibinfo{volume}{D85}},
  \bibinfo{pages}{116002} (\bibinfo{year}{2012}), \eprint{1109.2329}.

\bibitem[{ATL(2018)}]{ATLAS:2018yhd}
{``}\bibinfo{title}{{Search for supersymmetry in final states with missing
  transverse momentum and multiple $b$-jets in proton-proton collisions at
  $\sqrt{s} = 13$ TeV with the ATLAS detector}},{''} (\bibinfo{year}{2018}),
  \eprint{ATLAS-CONF-2018-041}.

\bibitem[{\citenamefont{Aad et~al.}(2020{\natexlab{b}})}]{ATLAS:2020xgt}
\bibinfo{author}{\bibfnamefont{G.}~\bibnamefont{Aad}} \bibnamefont{et~al.}
  (\bibinfo{collaboration}{ATLAS}), {``}\bibinfo{title}{{Search for new
  phenomena in final states with large jet multiplicities and missing
  transverse momentum using $ \sqrt{s} $ = 13 TeV proton-proton collisions
  recorded by ATLAS in Run 2 of the LHC}},{''} \bibinfo{journal}{JHEP}
  \textbf{\bibinfo{volume}{10}}, \bibinfo{pages}{062}
  (\bibinfo{year}{2020}{\natexlab{b}}), \eprint{2008.06032}.

\bibitem[{\citenamefont{Aaboud et~al.}(2017)}]{ATLAS:2017tmw}
\bibinfo{author}{\bibfnamefont{M.}~\bibnamefont{Aaboud}} \bibnamefont{et~al.}
  (\bibinfo{collaboration}{ATLAS}), {``}\bibinfo{title}{{Search for
  supersymmetry in final states with two same-sign or three leptons and jets
  using 36 fb$^{-1}$ of $\sqrt{s}=13$ TeV $pp$ collision data with the ATLAS
  detector}},{''} \bibinfo{journal}{JHEP} \textbf{\bibinfo{volume}{09}},
  \bibinfo{pages}{084} (\bibinfo{year}{2017}), \bibinfo{note}{[Erratum: JHEP
  08, 121 (2019)]}, \eprint{1706.03731}.

\bibitem[{\citenamefont{Sirunyan
  et~al.}(2020{\natexlab{a}})}]{Sirunyan:2019xwh}
\bibinfo{author}{\bibfnamefont{A.~M.} \bibnamefont{Sirunyan}}
  \bibnamefont{et~al.} (\bibinfo{collaboration}{CMS}),
  {``}\bibinfo{title}{{Searches for physics beyond the standard model with the
  $M_\mathrm{T2}$ variable in hadronic final states with and without
  disappearing tracks in proton-proton collisions at $\sqrt{s}=$ 13 TeV}},{''}
  \bibinfo{journal}{Eur. Phys. J. C} \textbf{\bibinfo{volume}{80}},
  \bibinfo{pages}{3} (\bibinfo{year}{2020}{\natexlab{a}}), \eprint{1909.03460}.

\bibitem[{\citenamefont{Sirunyan et~al.}(2019)}]{Sirunyan:2019ctn}
\bibinfo{author}{\bibfnamefont{A.~M.} \bibnamefont{Sirunyan}}
  \bibnamefont{et~al.} (\bibinfo{collaboration}{CMS}),
  {``}\bibinfo{title}{{Search for supersymmetry in proton-proton collisions at
  13 TeV in final states with jets and missing transverse momentum}},{''}
  \bibinfo{journal}{JHEP} \textbf{\bibinfo{volume}{10}}, \bibinfo{pages}{244}
  (\bibinfo{year}{2019}), \eprint{1908.04722}.

\bibitem[{\citenamefont{Sirunyan et~al.}(2021)}]{Sirunyan:2021mrs}
\bibinfo{author}{\bibfnamefont{A.~M.} \bibnamefont{Sirunyan}}
  \bibnamefont{et~al.} (\bibinfo{collaboration}{CMS}),
  {``}\bibinfo{title}{{Search for top squark production in fully-hadronic final
  states in proton-proton collisions at $\sqrt{s} =$ 13 TeV}},{''}
  (\bibinfo{year}{2021}), \eprint{2103.01290}.

\bibitem[{\citenamefont{Sirunyan et~al.}(2020{\natexlab{b}})}]{CMS:2019tlp}
\bibinfo{author}{\bibfnamefont{A.~M.} \bibnamefont{Sirunyan}}
  \bibnamefont{et~al.} (\bibinfo{collaboration}{CMS}),
  {``}\bibinfo{title}{{Search for supersymmetry in pp collisions at $\sqrt{s}=$
  13 TeV with 137 fb$^{-1}$ in final states with a single lepton using the sum
  of masses of large-radius jets}},{''} \bibinfo{journal}{Phys. Rev. D}
  \textbf{\bibinfo{volume}{101}}, \bibinfo{pages}{052010}
  (\bibinfo{year}{2020}{\natexlab{b}}), \eprint{1911.07558}.

\bibitem[{ATL(2021)}]{ATLAS_heavy}
{``}\bibinfo{title}{{Search for resonant and non-resonant Higgs boson pair
  production in the $b b^- \tau^+ \tau^-$ decay channel using 13 TeV pp
  collision data from the ATLAS detector}},{''} (\bibinfo{year}{2021}),
  \eprint{ATLAS-CONF-2021-030}.

\bibitem[{CMS(2021{\natexlab{b}})}]{CMS_heavy}
{``}\bibinfo{title}{{Search for new particles in an extended Higgs sector in
  the four b quark final state at $\sqrt{s} = 13$~TeV}},{''}
  (\bibinfo{year}{2021}{\natexlab{b}}), \eprint{CMS PAS B2G-20-003}.

\end{thebibliography}

\end{document}